\documentclass{jfm}
\usepackage{float}
\usepackage{subcaption}
\usepackage{caption}
\usepackage{amssymb}
\usepackage{newtxtext}
\usepackage{newtxmath}
\usepackage{hyperref}
\usepackage{gensymb}
\usepackage{svg}
\hypersetup{
colorlinks = true,
urlcolor = blue,
citecolor = black,
linkcolor = blue,
}
\usepackage{natbib}
\usepackage{authblk}

\newcommand{\RomanNumeralCaps}[1]
\title{A purely analytical and physical wind turbine wake model accounting for atmospheric stratification}

\author[1]{Emeline No\"el}
\author[1]{Erwan J\'ez\'equel}
\author[1]{Pierre-Antoine Joulin}
\affil[1]{IFP Energies nouvelles, 1 et 4 avenue de Bois-Préau, 92852 Rueil-Malmaison, France}

\begin{document}
\maketitle

\begin{abstract}
A purely analytical wake model for wind turbines is derived, anchored exclusively in physical interactions between atmospheric turbulence and turbine dynamics, and thus inherently accounting for atmospheric stratification. Unlike empirical models relying on assumed wake deficit shapes or tunable coefficients, this model predicts the wake deficit solely from measurable properties of the inflow—namely, turbulence intensity and the turbulence integral time scale. Systematic validation against Large Eddy Simulations (LES) for both IEA 15MW and NREL 5MW turbines—simulated in Meso-NH under stable, neutral, and unstable conditions—demonstrates excellent agreement across atmospheric regimes. Importantly, the model requires these specific turbulence statistics as input but shows only weak sensitivity to the integral time scale, ensuring robustness even with moderate uncertainties in inflow characterisation. Comparative analysis with the state-of-the-art Super-Gaussian analytical model highlights superior performance of the present approach, particularly for unstable and neutral stratification. These results show that the predictive accuracy gained by incorporating richer inflow physics justifies the need for more comprehensive atmospheric inputs, providing a clear pathway for physically grounded, calibration-free wake modeling in operational wind energy contexts.
\end{abstract}

\begin{keywords}
%Wind turbine wakes, Analytical wake model, Atmospheric stratification, Turbulence intensity, Integral time scale, Taylor diffusion, Large Eddy Simulation
\end{keywords}

\section{Introduction}\label{sec:intro}
%Wind power would be one of the largest sources of electricity worldwide according to IRENA's $1.5\degree$ scenario \citep{IRENA2023}.
Wind energy is expected to become a cornerstone of global electricity generation, with wind and solar projected to supply over 70\% of global power by 2050 under IRENA’s $1.5\degree$ scenario \citep{IRENA2023}. This projection highlights the critical role of renewable energy in the global energy transition, a perspective reinforced by international policy frameworks such as those discussed at COP28 \citep{COP282023}. To generate wind power, wind turbines are clustered together in a wind farm. As a result, many wind turbines operate in the wake of upstream turbines. A wind turbine's wake is a region of reduced wind speed and increased turbulence generated by the extraction of wind energy. 

\par
\begin{figure}
    \centering
    \includegraphics[width=\linewidth]{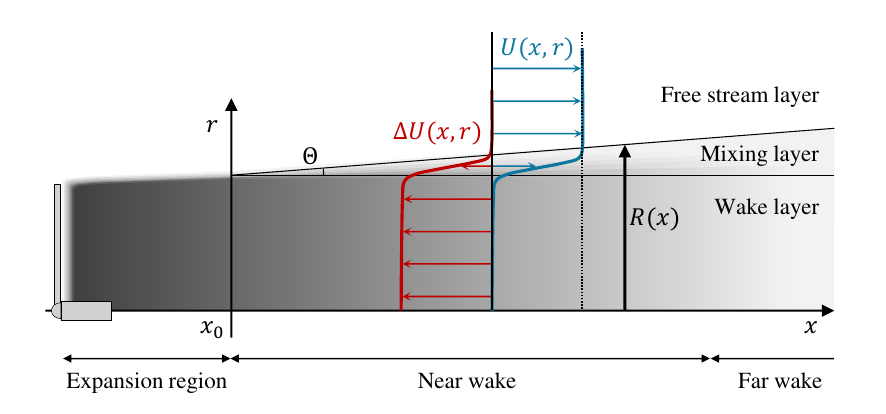}
     \caption{
Schematic diagram of a wind turbine wake illustrating the wake expansion angle $\Theta$. 
}
     \label{fig:tan}
 \end{figure}

 These wake effects reduce power generation and increase fatigue loads \citep{Frandsen1992, Barthelmie2010, Stevens2017, Porte-Agel2020}. Therefore, analytical models of the velocity deficit have been developed to optimise the design of wind farms for power generation \citep{Gocmen2016}. The self-similarity assumption suggests that the wake velocity deficit profile can be represented as the product of a velocity scale that varies with downstream distance and a normalized radial shape function. This relationship is generally expressed as:

\begin{equation}\label{Ushape}
\Delta U(x,r) = U_w(x)f\left(\frac{r}{R(x)}\right)
\end{equation}
Here, $U_w(x)$ denotes the maximum velocity deficit at position $x$ downstream (typically at the wake centerline), $r$ is the radial distance 
from the center of the wake, and $R(x)$ is the characteristic wake width at distance $x$. The function $f(r/R(x))$ describes the normalized 
velocity distribution across the wake. \cite{Bastankhah2014} proposed that the velocity distribution in the far wake region is well 
approximated by a Gaussian profile, as supported by wind tunnel experiments \citep{Chamorro2009} and earlier reviews \citep{Vermeer2003}. This 
led to the development of a widely used analytical wake model based on a Gaussian form for $f(r/R(x))$. Numerical studies have further 
corroborated the Gaussian nature of the velocity distribution in the far wake \citep{Xie2015}. Other velocity distributions, such as double 
Gaussian \citep{Keane2016} or super Gaussian \citep{Blondel2020}, have been proposed to unify near-wake and far-wake behaviours. However, the 
precise evolution of the characteristic wake width $R(x)$ downstream is not universally defined and generally requires empirical 
determination. To complete the analytical model, the wake width $R(x)$ is often assumed to increase linearly with downstream distance as $R(x) 
= \tan(\Theta)x/D + b$, where $\Theta$ is the wake expansion angle and $b$ is an initial offset. The value of $tan(\Theta)$ (the wake growth 
rate) must be determined from experimental or field data. Figure \ref{fig:tan} depicts a wind turbine wake and highlights how the wake growth rate is defined.
% \begin{figure}
%     \centering
%     \includegraphics[width=\linewidth]{images/fig1.png}
%      \caption{
% Schematic diagram of a wind turbine wake illustrating the wake expansion angle $\Theta$. The wake growth rate, given by $\tan(\Theta)$, is defined as the change in wake width from $R_1$ at $x_1$ to $R_2$ at $x_2$. This image was generated with assistance from the AI engine Illustrae (\url{https://illustrae.co})
% }
%     \label{fig:tan}
% \end{figure}
% \begin{figure}
%   \centering
%   \includesvg[width=0.75\textwidth]{fig1.svg}
%   \caption{Schematic diagram of a wind turbine wake illustrating the wake expansion angle $\Theta$.}
% \end{figure}
% \begin{figure}
%     \centering
%     \includegraphics[width=\linewidth]{images/fig1_bis.png}
%      \caption{
% Schematic diagram of a wind turbine wake illustrating the wake expansion angle $\Theta$.}
%     \label{fig:tan}
% \end{figure}
%Based on their numerical results for the Horns Rev wind farm, they established the following linear relationship:
%\begin{equation}\label{growRate}
%tan(\Theta) = 0.3837I_{u}+0.003678
%\end{equation}
\par\cite{Niayifar2016} proposed that the wake growth rate should depend directly on the streamwise ambient turbulence intensity $I_{u}$. While this relation emphasizes the role of turbulence intensity, it reflects a common feature of so-called bottom-up wake models: they typically represent the atmospheric boundary layer (ABL) through streamwise turbulence intensity alone, with limited focus on its vertical structure or dynamic behaviour. In contrast, top-down or single-column models represent the wind farm as an internal boundary layer within the ABL, capturing the collective impact of turbines on the vertical wind profile under the assumption of an infinite, uniformly spaced turbine array \citep{Frandsen1992, Calaf2010}. These models account for key ABL features, such as surface roughness, Coriolis force, large-scale pressure gradients, and atmospheric stability \citep{Pena2014}. Notably, they have shown that strong thermal stratification in the free atmosphere can lead to shallower boundary layer height. This limits the entrainment of kinetic energy from the overlying flow, ultimately decreasing farm power output. 
\par Couplings between top-down and bottom-up models offer a physically grounded alternative to purely empirical approaches by accounting for atmospheric effects on wake growth rate. For instance, \cite{Pena2014} introduced a one-way coupling between the \cite{Frandsen1992} model and the \cite{Jensen1983} wake model to adjust the wake growth rate coefficient. More recently, \cite{Stevens2016} developed the Coupled Wake Boundary Layer (CWBL) model, which implements an iterative two-way coupling between the top-down model of \cite{Calaf2010} and the bottom-up model of \cite{Jensen1983}.  By incorporating effective wake coverage, this approach relaxes the infinite-farm assumption, enabling realistic modeling of finite and non-uniform turbine layouts. Complementary to these couplings, wake-added turbulence (WAT) models offer a practical way to account for atmospheric effects within bottom-up frameworks, such as limited ABL height, without resolving its detailed structure (\cite{Crespo1996, Frandsen2007, Ishihara2018}).  Drawing on the physical role of turbulence in wake recovery, these models are primarily empirical, relying on calibration with ambient turbulence intensity (\cite{Niayifar2016, Blondel2020a, Pena2014}). Recent advancements have shown that combining WAT models with a super-Gaussian wake model and a momentum-conserving superposition approach leads to strong agreement with large eddy simulation (LES) results, as demonstrated by \cite{Blondel2023} for the Horns Rev 1 wind farm. 
\par However, streamwise turbulence intensity alone does not fully capture the influence of atmospheric stability on wake recovery. Large eddy simulations performed by \cite{Du2021} reveal that, even with identical turbulence intensity, variations in atmospheric stratification alter turbulence anisotropy and spatial distribution. Convective conditions, in particular, enhance spanwise turbulent fluctuations, accelerating wake recovery. These findings align with earlier LES results by \cite{Xie2017}, which reported lower power deficits under unstable conditions, and with LIDAR-based field observations by \cite{Iungo2014}. Together, these studies underscore the critical role of turbulence structure—beyond turbulent intensity—in shaping wake dynamics.
\par The pioneering work of \cite{Ainslie1988} noted that wake deficits observed in the field were smaller than those measured in wind tunnel experiments. This difference was attributed to wind direction variability, leading to the introduction of a correction for the centreline velocity deficit based on the standard deviation of wind direction. Building on this, \cite{Larsen2008} linked wake meandering to large-scale atmospheric turbulence, proposing that the wake behaves as a passive scalar advected by these turbulent structures. This concept underpins the Dynamic Wake Meandering (DWM) model, which uses a stochastic velocity field to transport the wake structure and enables simultaneous assessment of turbine performance and loading, particularly intermittent loads. In parallel, the passive scalar transport assumption has also supported the development of physically based wake growth rate models \citep{Cheng2018}, grounded in Taylor’s diffusion theory \citep{Taylor1922}. Taylor’s diffusion theory explains how random velocity fluctuations in turbulence lead to the diffusive spreading of a passive scalar over long timescales. The effective diffusivity is determined by the velocity autocorrelation of the flow. While these models perform well under neutral conditions, their accuracy is deteriorated in low-turbulence or strongly stratified regimes, especially when turbine-induced turbulence is not considered. Recent improvements include the explicit incorporation of turbine-induced turbulence \citep{Vahidi2022} and extension to a wider range of atmospheric boundary layer (ABL) stability conditions \citep{Du2022}. More recently, \cite{Ali2024} developed an analytical wake model based on the passive scalar analogy, directly solving the diffusion equation for a circular disk source. This approach yields a closed-form solution for the concentration profile, which is used to represent the wake velocity deficit distribution and ensures conservation of linear momentum. However, the model still relies on empirical specification of the wake growth rate and requires distinguishing between the near-wake and far-wake regions. Furthermore, the model does not explicitly account for different atmospheric stability conditions, as the wake growth rate is parameterized solely by the ambient turbulence intensity, without directly considering the effects of atmospheric stratification.
\par Bridging the stochastic wake advection of the DWM and the diffusive scalar perspective, \cite{Braunbehrens2019} proposed a unified framework by modeling the wake deficit as the convolution of the probability distribution of the wake center position with an underlying velocity deficit distribution. This formulation captures both the displacement effects of large-scale turbulence and the local structure of the wake. \cite{Jezequel2024a} extended this perspective by explicitly linking wake transport to coherent atmospheric structures, confirming through LES that the wake deficit behaves like a passive scalar across different stability regimes. They also generalized the approach to turbulence intensity \citep{Jezequel2024b}, opening new pathways for physically grounded wake modeling. However, these frameworks present an inconsistency in applying Taylor’s diffusion theory to describe the movement of the wake center, which is not a single particle but rather the spatial average of fluid parcels within the wake. \cite{Noel2025} addressed this issue by providing a theoretical framework that justifies the use of Taylor’s diffusion theory for wake center dispersion, showing how it can be rigorously derived using Eulerian velocity statistics. This approach unifies the Eulerian description of wake meandering with the Lagrangian interpretation of wake growth, highlighting how both phenomena emerge from the underlying turbulence and jointly shape wind turbine wake behaviour. 
\par Despite these advances, several challenges remain. Most existing models —including recent analytical approaches— continue to depend on assumed forms for the normalized velocity deficit distribution (such as Gaussian or super-Gaussian profiles) or require empirical calibration, and often involve iterative solution procedures. Moreover, their accuracy can be limited under convective or strongly stable atmospheric stratification. To date, no model offers a fully analytical framework capable of predicting both the wake velocity deficit distribution and its downstream evolution based solely on physical phenomena. This highlights the ongoing need for new modeling strategies that can robustly and accurately capture wind turbine wake behaviour across a wide range of atmospheric conditions.

\par This study seeks to overcome these limitations by removing the need to prescribe a velocity deficit distribution or perform empirical calibration. Instead, we propose a fully analytical, physically based model that predicts the wake velocity deficit using only the upstream atmospheric flow properties. This approach aims to enhance model accuracy, generalisability, and ease of use, while providing a clearer physical understanding of turbine–atmosphere interactions. To do so, we build upon prior studies, particularly those linking atmospheric turbulence to wake dispersion, and draw on the concept of the wind turbine acting as a low-pass filter of the incoming flow. The structure of the paper is as follows:  Section \ref{WakeDerivation} details the derivation of our analytical model and its theoretical underpinnings, Section \ref{ModelValidation} presents a validation of the model against high-fidelity Large Eddy Simulation (LES) data, demonstrating its predictive capabilities. Finally, Section \ref{Conclusion} concludes with a discussion of the key insights, implications, and opportunities for future work.

\section{Wake model derivation}\label{WakeDerivation}
This section details the derivation of the model by analysing fluid parcel behaviour through probability addition. The initial ensemble of fluid parcels that constitute the wake is defined by the probability of their presence within the rotor area. These parcels, treated as passive scalars, are displaced in the same manner as those in the ambient turbulent flow. Their displacement is described using Taylor’s diffusion theory. Beyond this passive scalar behaviour, the interaction between the wake and the surrounding flow generates a mixing layer, whose growth downstream is also incorporated with a mixing layer growth model. The velocity deficit shape function is then given by the convolution of the initial probability density function and the displacement probability distribution. To evaluate the norm of the velocity deficit, the energy dissipation is quantified by applying the cut-off frequency of fluid displacement to determine the damping of the L2 norm of the wake velocity deficit. This methodology ultimately provides analytical expressions for both the maximum velocity deficit and the corresponding shape function.

\subsection{Turbine : initial wake structure probability density function}

Wake meandering behind turbines manifests as coherent flow structures \citep{Medici2007}, analogous to vortex shedding behind circular bluff bodies \citep{Miau1997}, with a characteristic shedding frequency expressed by a Strouhal number. Both upstream velocity fluctuations and the wake center position exhibit strong spectral coherence at this frequency \citep{Mao2018,DeCillis2022}. Alternatively, \citet{Larsen2008} interpret the spatial extent of the rotor as imposing a cut-off frequency on incoming flow fluctuations, thus explaining wake meandering via spatial filtering. It is proposed to reconciled these viewpoints by considering the shedding frequency as a cut-off frequency, consistent with the turbine functioning as a low-pass filter that attenuates high-frequency, small-scale atmospheric disturbances.

Therefore, given that the turbine acts as a low-pass filter on the atmospheric flow, it is modeled here as an ideal low-pass filter. To simplify the analysis, variations are considered only along the angular direction, effectively reducing the problem to one dimension. The ideal low-pass filter is defined as: %The rotor disc’s axisymmetry permits reduction of the problem from two dimensions to one, by focusing on a single radial cross-section.

\begin{equation}
H(s_D) =
\begin{cases}
1, & \text{for } -0.5 \leq s_D \leq 0.5 \\
0, & \text{otherwise}
\end{cases}
\label{gat}
\end{equation}

where $s_{D}$ denotes a coordinate in an arbitrary direction across the rotor disk, normalized by the turbine diameter. The rectangular function defined by equation \eqref{gat} can be interpreted as a spatial probability distribution describing the likelihood of encountering the rotor disc at position $s_D$. This distribution thus defines the initial spatial structure imposed by the turbine on the incoming flow

\subsection{Turbulence : fluid parcels displacement probability density function}
\subsubsection{Mixing layer : turbine-induced displacement}

As \citet{Cheng2018} and \citet{Du2022} report, besides the filtering effect of the turbine on large atmospheric scales, some scales result from the direct interaction between the wake structure deficit and the free-flow. These additional turbulent scales are important to consider in very stable atmospheric conditions or low turbulence scenarios, as the scales of atmospheric turbulence tend to be smaller. \citet{Vahidi2022} proposed that the modelling of this turbine-induced effect should be approached as a turbulent mixing layer between the wake layer and the free-stream layer. According to \citet{Pope2000}, a mixing layer grows linearly with distance downstream. The linear constant, known as the spreading parameter, can be expressed as follows:

\begin{equation}
S =\frac{U_c}{U_s}\frac{d\sigma_{ML}}{dx} 
\label{ML}
\end{equation}

$\sigma_{ML}$ is the characteristic length of the mixing layer, defined as the displacement of a fluid parcel across the mixing layer along a radial cross-section. $U_s$ is the characteristic shear velocity \eqref{Us} and $U_c$ is the convective velocity \eqref{Uc}. Considering the \citet{Pope2000}'s description of the mixing layer involving two uniform, parallel flows, it can be argued that the amplitude of wake velocity deficit provides a more representative measure of the velocity within the wake layer than the centreline velocity. Consequently, the shear and convective layer velocities are here defined using the L1 norm of the wake velocity deficit, rather than the centreline velocity as employed by \citet{Vahidi2022}. 

\begin{equation}
U_s = \| \Delta U(x,s_D)\|_{\text{1}}
\label{Us}
\end{equation}
\begin{equation}
U_c = U_{\infty}-\frac{\|  \Delta U(x,s_D) \|_{\text{1}}}{2}
\label{Uc}
\end{equation}

Accordingly, the equation \eqref{ML} can be expressed as follows:

\begin{equation}
U_{c}\frac{d\sigma_{ML}}{dx} = 2(U_{\infty}-U_{c})S
\label{MLx}
\end{equation}

The initial point of wake development, designated as $x_0$, is presumed to be situated one rotor diameter downstream of the turbine, as in \citet{Vahidi2022}. In the expansion region, prior to the initial point of wake development ($x_0$), two key phenomena occur: the pressure recovers to ambient levels, and the growth of the shear layer leads to a small layer compared to the rotor diameter. Beyond this point, the wake begins to develop in a more pronounced manner. As a result, to determine the characteristic length at a downstream position, the equation is integrated from this initial point of wake development ($x_0$) to the considered position ($x$). This approach allows to account for the wake evolution from the point where it truly begins to develop, thus excluding the initial expansion region where changes are less significant. The integration yields:

\begin{equation}
\sigma_{ML}(x) =  2S\left[\int_{x_0}^x \frac{U_{\infty}}{U_{c}}dx-\int_{x_0}^x  dx\right]=  2S(U_{\infty}T-(x-x_0))
\label{MLx}
\end{equation}

As suggested by \citet{Vahidi2022}, the spreading parameter $S$ is set to $0.043$. $T$ is defined as a characteristic time expressed in terms of the convective velocity of the wake:

\begin{equation}
T = \int_{x_0}^x \frac{1}{U_{c}}dx
\label{Travel}
\end{equation}

\subsubsection{Taylor diffusion : atmospheric-induced displacement}

In his seminal work, \citet{Taylor1922} introduced a groundbreaking framework for modelling the influence of turbulence on scalar quantities by treating turbulent transport as a diffusion process analogous to molecular diffusion. The central insight of Taylor’s analysis is that, despite the apparent randomness of turbulence, the motion of fluid parcels is governed by statistical correlations over time. These correlations, quantified by the velocity auto-correlation function, determine the effective diffusive transport in turbulent flows. The auto-correlation function is defined as:

\begin{equation}
\label{autoCorr}
R_{u_i}(\tau) = \lim_{T \to \infty} \frac{1}{T} \int_0^T u_i(t) \, u_i(t+\tau) \, dt
\end{equation}
where $u_i$ denotes the $i$-th component of velocity and $\tau$ is the time lag. The integral time scale $A_i$, which quantifies the timescale over which velocity correlations decay, is then defined as:

\begin{equation}
A_i = \frac{1}{R_{u_i}(0)} \int_0^{+\infty} R_{u_i}(\tau) \, d\tau
\label{Auto}
\end{equation}
This time scale represents a characteristic measure of the memory of the flow, capturing the timescale over which the velocity fluctuations remain correlated.

%In the Dynamic Wake Meandering (DWM) framework, the wake is assimilated to a scalar that is passively advected downstream by turbulence \citep{Larsen2008}. Taylor's theory has therefore been applied to meandering \citep{Braunbehrens2019} to provide an estimation of its characteristic displacement. Assuming an exponential decay of the velocity auto-correlation, the characteristic displacement of a fluid parcel in direction $i$ (where $i = v$ for lateral or $i = w$ for vertical) is given by Taylor’s formulation:

Considering the wake structure as a passive scalar advected by turbulence \citep{Larsen2008,Jezequel2024a}, the displacement of a fluid 
parcel within the wake can be described using Taylor's diffusion theory. Assuming an exponential decay of the velocity auto-correlation, the characteristic displacement of a fluid parcel in direction $i$ (where $i = v$ for 
lateral or $i = w$ for vertical) is given by Taylor’s formulation:
\begin{equation}
\sigma_{{LS}_i}(x) = \sigma_i \sqrt{2A_iT - 2A_i^2\left(1 - e^{-T/A_i}\right)}
\label{LS}
\end{equation}
where $\sigma_{{LS}_i}(x)$ represents the characteristic lateral or vertical path length of a fluid parcel within the wake, resulting from large-scale atmospheric turbulence. Here, $T$ is defined by equation \eqref{Travel}, $\sigma_i$ is the standard deviation of the lateral or vertical components of the inflow velocity, and the integral time scale $A_i$ is given by:

\begin{equation}
A_i = \frac{1}{R_{u_i}(0)} \int_0^{+\infty} R_{u_i}(\tau) \, d\tau
\label{Auto}
\end{equation}

It is noteworthy that the integral time scale referenced by \citet{Taylor1922} represents the integral time scale of fluid parcel displacement, which is equivalent to the Lagrangian time scale. Nevertheless, it is typically more straightforward to work with the Eulerian time scale, as it can be calculated using the fluid velocity time series.  As suggested by \citet{Hay1959}, the ratio $\gamma$ between the Lagrangian and Eulerian time scales ($A_i^L$ and $A_i^E$ respectively) is assumed to be proportional to the inverse of the turbulence intensity $I_i$ as follows: 

\begin{equation}
\frac{A_i^L}{A_i^E}= \frac{\gamma}{I_i}
\label{Ratio}
\end{equation}

\citet{Anfossi2006} have investigated this ratio using LES simulations. In particular, they provide some estimates of $\gamma$ for the convective ABL ($\gamma \sim 0.6$) and for the neutral ABL ($\gamma \sim 0.4$). These estimations are employed in order to achieve a conversion from the Eulerian time scale to the Lagrangian time scale.

\subsubsection{Fluid parcel radial cross-section displacement probability density function}
% Note : characterize by its path length
The total displacement of a fluid parcel within a wake at a downstream location can be understood as a combination of two key components. The first component occurs within the wake's mixing layer as a consequence of turbulent mixing processes. This phenomenon represents the local movement of fluid parcels within the wake structure itself. The second component affects the entire wake structure. This phenomenon is caused by broader atmospheric flow dynamics that act on the wake structure as a whole. The total displacement of a fluid parcel within a wake, referred to as the characteristic path length, is then determined by superimposing these two components. This superposition reflects both the internal turbulent dynamics of the wake and the external influences of the atmosphere. This can be expressed as follows:

\begin{equation}
\sigma_{all_i}(x) = \underbrace{\frac{\sigma_{ML}(x)}{D}}_{\text{internal mixing}}+\underbrace{\frac{\sigma_{{LS}_i}(x)}{D}}_{\text{external mixing}}
\label{SumDisy}
\end{equation}

$\sigma_{all_i}$ is the normalized characteristic path length, where $i$ denotes either $y$ (lateral) or $z$ (vertical), at location $x$. In order to reduce the dimensionality of the problem, it is assumed that the wake is axisymmetric, which is a common assumption \citep{Bastankhah2014,Blondel2019,Frandsen2006,Cheng2018}. It follows that an appropriate combination of the characteristic path lengths is essential. In line with \citet{Vahidi2022}, the geometric mean is employed: $\sigma_{all}(x) = \sqrt{\sigma_{all_y}(x)\sigma_{all_z}(x)}$.

Under the ergodic hypothesis, the characteristic path length $\sigma_{all}(x)$ serves as an indicator of the turbulence scale and reflects the statistical properties of the observed fluid parcel displacements of the wake. These displacements should be interpreted as deviations from the mean motion of the wake structure. Consequently, this characteristic path length can be associated with a standard deviation that quantifies how much individual displacements differ from this mean motion. According to the central limit theorem, if a sufficiently large number of independent observations of the displacement of fluid parcels at a given location are examined, their displacement distribution can be approximated by a normal distribution with zero mean and the characteristic path length acting as the standard deviation. Therefore, the probability of fluid parcel displacement at location $x$, would be expressed as follows: 

\begin{equation}
f_x(s_{D}) = \frac{1}{\sigma_{all}(x)\sqrt{2\pi}}e^{-\frac{{s_D}^2}{2\sigma_{all}(x)^2}}
\label{SumDisy}
\end{equation}

Here, $s_{D}$ denotes a coordinate in an arbitrary direction across the rotor disk, normalised by the turbine diameter.
%If these displacements are regarded as independent, the resulting path length can be expressed as the Euclidean norm $\sigma(x) = \sqrt{\sigma_y(x)^2+\sigma_z(x)^2}$. An alternative approach is to view these displacements as dependent. In this case, the resulting path length would be represented by the geometric mean $\sigma(x) = \sqrt{\sigma_y(x)\sigma_z(x)}$. In situations where the lateral standard deviation of velocity exceeds the axial standard deviation, it can be concluded that the lateral or vertical turbulent components are not related to streamwise turbulence. This suggests that the lateral and vertical components behave independently. Conversely, when the axial standard deviation of velocity is greater than the lateral standard deviation, it indicates that all turbulent components are interconnected, including those in the lateral and vertical directions. Accordingly, the ratio of standard deviations, namely, $\sigma_v/\sigma_u$, is proposed as an indicator of the path's independence behavior.
\subsection{Probability density function of downstream wake's spatial distribution}

The downstream statistical evolution of the wake can be described as the result of two key probabilistic processes. First, fluid parcels from the incoming flow intersect the turbine’s rotor disc, forming the initial wake structure. This event is represented by the probability density function given in equation \eqref{gat}. As the wake propagates downstream, turbulent motions displace fluid parcels from their initial positions. At a given downstream location ($x$), the probability density for the displacement of a fluid parcel is given by equation \eqref{SumDisy}. The overall probability of finding a fluid parcel at a particular position in the wake, accounting for both its initial location at the rotor disc and its subsequent downstream displacement, is given by the convolution of these two probability density functions. This convolution effectively combines the initial wake structure with its downstream evolution due to turbulence, providing a comprehensive probabilistic description of the wake’s spatial distribution at any given downstream distance. This convolution can be expressed analytically in terms of the error function as follows:

\begin{equation}
W_x(s_D) = [H * f_x](s_D) = \frac{1}{2} \left[\text{erf}\left(\frac{s_D + 0.5}{\sqrt{2}\sigma_{all}(x)}\right) - \text{erf}\left(\frac{s_D - 0.5}{\sqrt{2}\sigma_{all}(x)}\right)\right]
\label{Wakepdf}
\end{equation}
\\
where $W_x(s_D)$ denotes the probability density function (PDF) describing the wake's spatial distribution at location ($x$), and $s_D$ represents a normalized coordinate in an arbitrary direction across the rotor disk.

\subsection{Normalisation factor for the probability function of the wake's spatial distribution}

The objective of this study is to develop a statistical model for predicting the distribution of wake velocity deficits. To this end, we derive a PDF characterising the spatial distribution of the wake. However, this PDF does not directly represent the wake velocity deficit magnitude. To obtain a quantitative description of the velocity deficit distribution, an appropriate normalisation factor must be determined. Accordingly, the wake velocity deficit distribution at a given streamwise location ($x$) is expressed as

\begin{equation}
\Delta U(x,s_D) = \alpha_x W_x(s_D)
\label{WD}
\end{equation}
\\
where $\alpha_x$ denotes the normalisation factor at location $x$. To evaluate $\alpha_x$, we use the energy content of the wake velocity deficit, which is quantified by the $L^2$ norm. The normalisation factor is then given by:

\begin{equation}
 \alpha_x= \frac{\|\Delta U(x,s_D)\|_{\text{2}}}{\|W_x(s_D)\|_{\text{2}}}
\label{WD_L2}
\end{equation}
\\

In the Dynamic Wake Meandering (DWM) framework, the transformation from the moving to the fixed reference frame for the wake velocity deficit is performed by convolving the wake center probability density function (PDF) with the velocity deficit in the moving frame \citep{Keck2014,Braunbehrens2019,Jezequel2024b}. As demonstrated by \citet{Jezequel2024a}, this approach introduces negligible error across a range of atmospheric stability conditions. This methodology thus provides a robust basis for expressing the distribution of the wake velocity deficit as a function of the initial wake velocity deficit $U_0$, as follows:

%The mean velocity in the fixed frame of reference can be computed as the convolution of the wake center PDF with the mean velocity in the moving frame of reference with minimal error \citep{Jezequel2024a}. This approach, commonly used in the Dynamic Wake Meandering framework \citep{Keck2014,Braunbehrens2019,Jezequel2024b},  provides a valuable foundation for expressing the distribution of the wake velocity deficit as a function of the initial wake velocity deficit $U_0$ as follows:

\begin{equation}
\Delta U(x,s_D) = [U_0*f_x](s_D)
\label{Uo}
\end{equation}
\\
In accordance with the Young's convolution inequality \citep{Saitoh2000}, the L2 norm of the wake velocity deficit can be bounded as follows:
\begin{equation}
\|\Delta U(x,s_D)\|_{\text{2}}\hspace{2ex} \leq \hspace{2ex} \|U_0\|_{\text{2}} \hspace{2ex} \|f_x(s_D)\|_{\text{1}}
\label{Norm}
\end{equation}
\\
Using the 1D momentum theory for an ideal wind turbine \citep{Hansen1998}, we expressed the L2 norm of the initial wake velocity deficit, the previous equation can be rewritten as:
\begin{equation}
\alpha_x \|W_x(s_D)\|_{\text{2}}\hspace{2ex} \leq \hspace{2ex} U_\infty(1-\sqrt{1-C_T}) \hspace{2ex}\|f_x(s_D)\|_{\text{1}}
\label{Norm2}
\end{equation} 
It is reasonable to propose that the fluid parcel displacement, modelled as a Gaussian PDF, serves as a filter for the initial wake deficit velocity. The cut-off frequency, which corresponds to the half-power point, can be expressed as $s_{D_c} = \sqrt{2 \ln(2)} \sigma(x_0)$ \citep{Bottacchi2008}. Our assumption is that displacements outside this bandwidth would decrease the energy of the wake deficit velocity. Accordingly, it is assumed that the bounding of the L2 norm of the wake velocity deficit may be approximated by the following expression:

\begin{equation}
\alpha_x \|W_x(s_D)\|_{\text{2}}\hspace{2ex} = \hspace{2ex}U_\infty(1-\sqrt{1-C_T}) \hspace{2ex} \int_{-s_{D_c}}^{s_{D_c}}f_x(s_D)ds_D 
\label{Norm2}
\end{equation} 

which can be simplified as:

\begin{equation}
\alpha_x \|W_x(s_D)\|_{\text{2}}\hspace{2ex} =\hspace{2ex}U_\infty(1-\sqrt{1-C_T}) \hspace{1ex}\text{erf}\left(\frac{s_{D_c}}{\sigma_{all}(x)\sqrt{2}}\right) 
\label{Norm2}
\end{equation} 

Therefore, the normalisation constant can be expressed as :

\begin{equation}
\alpha_x \hspace{2ex} =\frac{\hspace{2ex}(1-\sqrt{1-C_T})U_\infty \hspace{1ex}\text{erf}\left(\frac{s_{D_c}}{\sigma_{all}(x)\sqrt{2}}\right)}{\|W_x(s_D)\|_{\text{2}}} 
\label{Norm2_vf}
\end{equation} 
\newline

Considering the parity of the error function, the square of the L2 norm of the wake's spatial PDF is expressed as follows:

\begin{equation}
\int_{-\infty}^{\infty} W_x(s_D)^2 \, ds_D = \int_{-\infty}^{\infty}\left[\frac{1}{2} \left[\text{erf}(a+g(s_D)) + \text{erf}(a-g(s_D))\right]\right]^2\, ds_D
\label{L2W_parity}
\end{equation}
where
\begin{equation}
a = \frac{1}{2\sigma_{all}(x)\sqrt{2}} \quad \text{and} \quad g(s_D) = \frac{s_D}{\sigma_{all}(x)\sqrt{2}}
\label{L2W_parits_Def}
\end{equation}
\newline \\
The equation \eqref{L2W_parity} can be expanded as:

\begin{equation}
\begin{split}
\int_{-\infty}^{\infty} W_x(s_D)^2 \, ds_D &= \int_{-\infty}^{\infty}\frac{1}{4} \left[\text{erf}^2(a+g(s_D)) + \text{erf}^2(a-g(s_D))  \right]\, ds_D \\
& + \hspace{1ex} \int_{-\infty}^{\infty}  \left[\frac{1}{2} \hspace{0.5ex}\text{erf}(a+g(s_D))\text{erf}(a-g(s_D))\right] \, ds_D
\end{split}
\label{L2W_exp}
\end{equation}

Using the approximation $\text{erf}^2(x) \approx 1 - e^{-\xi^2x^2}$ with $\xi = 1.1131$, it is simplified as:

\begin{equation}
\begin{split}
\int_{-\infty}^{\infty} W_x(s_D)^2 \, ds_D &= \int_{-\infty}^{\infty}\frac{1}{2} \left[1 + \text{erf}(a+g(s_D))\text{erf}(a-g(s_D))\right]  \, ds_D \\
&-\int_{-\infty}^{\infty} \frac{1}{4}e^{-\xi^2(a+g(s_D))^2}\, ds_D - \int_{-\infty}^{\infty}\frac{1}{4}e^{-\xi^2(a-g(s_D))^2} \, ds_D
\end{split}
\label{eq:L2W_approximation}
\end{equation}

After applying function composition integration and using the properties of Gaussian integrals, this expression reduces to:

\begin{equation}
\int_{-\infty}^{\infty} W_x(s_D)^2 \, ds_D = \sigma_{all}(x)\sqrt{2}\left(2a\,\text{erf}(\sqrt{2}a) + \frac{\sqrt{2}}{\sqrt{\pi}}e^{-2a^2}\right) - \frac{1}{2}\sqrt{\frac{\pi}{\beta}}
\label{eq:L2W_final}
\end{equation}
where
\begin{equation}
\beta = \frac{\xi^2}{2\sigma_{all}^2(x)}
\label{beta}
\end{equation}

Finally, the L2 norm of $W_x(s_D)$ can be expressed as follows:

\begin{equation}
\|W_x(s_D)\|_{\text{2}} = \sqrt{\sigma_{all}(x)\sqrt{2}\left(2a\,\text{erf}(\sqrt{2}a) + \frac{\sqrt{2}}{\sqrt{\pi}}e^{-2a^2}\right) - \frac{1}{2}\sqrt{\frac{\pi}{\beta}}}
\label{Idef}
\end{equation} 

Then, the normalisation factor is written :

\begin{equation}
\alpha_x \hspace{2ex} =\frac{\hspace{2ex}(1-\sqrt{1-C_T})U_\infty \hspace{1ex}\text{erf}\left(\frac{s_{D_c}}{\sigma_{all}(x)\sqrt{2}}\right)}{\sqrt{\sigma_{all}(x)\sqrt{2}\left(2a\,\text{erf}(\sqrt{2}a) + \frac{\sqrt{2}}{\sqrt{\pi}}e^{-2a^2}\right) - \frac{1}{2}\sqrt{\frac{\pi}{\beta}}}} 
\label{Normalize}
\end{equation} 

\subsection{Downstream wake's spatial distribution}

The wake velocity deficit distribution at location ($x$) is expressed using the equation \eqref{WD} and the equation \eqref{Normalize} as:

\begin{align}
\Delta U(x,s_D) = & \frac{(1-\sqrt{1-C_T})U_\infty \,\text{erf}\left(\frac{s_{D_c}}{\sigma_{all}(x)\sqrt{2}}\right)}
                 {\sqrt{\sigma_{all}(x)\sqrt{2}\left(2a\,\text{erf}(\sqrt{2}a) + \frac{\sqrt{2}}{\sqrt{\pi}}e^{-2a^2}\right) - \frac{1}{2}\sqrt{\frac{\pi}{\beta}}}} \nonumber \\
           & \times \frac{1}{2} \left[\text{erf}\left(\frac{s_D + 0.5}{\sqrt{2}\sigma_{all}(x)}\right) - \text{erf}\left(\frac{s_D - 0.5}{\sqrt{2}\sigma_{all}(x)}\right)\right]
\label{eq:Analytic}
\end{align}
where
\begin{equation}
\sigma_{all}(x) = \sqrt{\sigma_{all_y}(x)\sigma_{all_z}(x)}
\label{geomean}
\end{equation}
The characteristic path length $\sigma_{all}(x)$ depends on the characteristic time $T$ defined in equation \eqref{Travel}, which can be written as:
\begin{equation}
T = \int_{x_0}^x \frac{1}{U_{c}} dx = \int_{x_0}^x \frac{1}{U_{\infty}-\frac{\alpha_x}{2}} dx
\label{Travel}
\end{equation}

However, $U_{\infty} - \alpha_x/2$ represents the mean wake velocity at location $x$, which already incorporates the cumulative effects of wake evolution due to diffusion from $x_0$ to $x$. In other words, this mean velocity at $x$ is itself the result of the integrated wake dynamics up to that point. Therefore, the characteristic time can be directly expressed as:
\begin{equation}
T(x) = \frac{x - x_0}{U_{\infty} - \frac{\alpha_x}{2}}, \quad \text{where } x_0 = 1D.
\label{Travel2}
\end{equation}
%It is assumed that the convective velocity \eqref{Uc} at the specific downstream location ($x$) serves as an accurate estimation of the velocity of fluid parcels travelling to that location. Consequently, one can express the travel time as:

%The minimum characteristic time is calculated using the one-dimensional momentum theory. Consequently, the characteristic time at a downstream location can be expressed as follows:

%\begin{equation}
%T(x) = \max\left(\frac{D}{0.5 U_{\infty} \left(1 + \sqrt{1 - C_T}\right)}, %\frac{x - x_0}{U_{\infty} - \frac{\alpha_x}{2}}\right)
%\label{Travel3}
%\end{equation}
Given the coupling between characteristic path length and characteristic time, the system can be treated as coupled. In practice, a fixed-point iteration may be used. Alternatively, a simpler method is to first estimate the characteristic time using the convective velocity from one-dimensional momentum theory, calculate the first characteristic path length, update the characteristic time, then obtain the final characteristic path length.

\subsection{Conclusion on the derivated wake model}

The wake model presented here provides a physically-based alternative to conventional approaches that rely on prescribed velocity deficit profiles. By explicitly linking atmospheric inflow characteristics with wind turbine response, the model enables wake predictions that are adaptable to a wide range of atmospheric conditions. The framework requires as inputs the key atmospheric parameters—turbulence root mean square velocities ($\sigma_u$, $\sigma_v$, $\sigma_w$), Lagrangian integral time scales ($A_v^L$, $A_w^L$), and streamwise mean velocity ($U_{\infty}$)—as well as standard turbine properties such as rotor diameter ($D$) and thrust coefficient ($C_T$). The implementation steps are detailed in Appendix \ref{Tech}.

\section{Model Validation}\label{ModelValidation}

This section details the validation of the model by comparing its predictions with LES. The main equations solved are presented, including the anelastic approximation and the turbulence modelling approach. The simulated case setups are described, covering three different atmospheric types. Results from these simulations, along with data from the literature, are then reviewed to provide an overview of the current model’s performance across various stability regimes and turbulence conditions.

\subsection{LES solver}

In order to validate the derived model and evaluate its performance in a stratified atmospheric boundary layer, large eddy simulations (LES) were conducted using the Meso-NH solver. Meso-NH is an open-source, non-hydrostatic meso- to micro-scale atmospheric model developed collaboratively through the joint efforts of the Laboratoire d'Aérologie and the Centre National de Recherches Météorologiques (CNRM) \citep{Lac2018}. The model is highly adaptable, capable of simulating both real-world scenarios and academic cases, and emulates various modelling approaches, including mesoscale meteorological models, cloud-resolving models (CRM), and LES models. \citet{Joulin2019} has introduced actuator methods to effectively model wind turbines within this framework. The solver implements the anelastic approximation equations, with a particular focus on Durran's pseudo-incompressible approximation (PIA) \citep{Durran1989, Durran2008}. Similar to low-Mach number solvers, this system filters out acoustic waves, thereby enabling efficient simulations of low-speed flows where sound waves are not critical. In this formulation, pressure acts as a Lagrange multiplier to enforce mass conservation, leading to a modified continuity equation that effectively eliminates high-frequency acoustic waves while accurately capturing significant atmospheric motions such as convection and gravity waves \citep{ACHATZ2010}. The system is derived for perfect gases using the Exner function, which retains the non-linearised equation of state. The Exner function is defined as:

\begin{equation}
\Pi = \left(\frac{P}{P_o}\right)^{\frac{R}{C_p}} 
\label{Exner}
\end{equation}
where $P$ is the thermodynamical pressure, $C_p$ is
the specific heat at constant pressure, $R$ is the gas constant, $P_o$ is a constant reference pressure. This approach maintains full thermodynamic relationships, providing more accurate representations of atmospheric processes. To formulate the PIA, a reference state denoted by $\overline{\phi(z)}$ — representing the hydrostatically balanced background atmosphere varying only with height $z$ — is introduced. Any variable $\phi$ is decomposed into this reference state plus a fluctuation:
\begin{equation}
\phi(x,y,z,t) = \overline{\phi(z)} + \phi'(x,y,z,t).
\end{equation}
Furthermore, within the LES framework, variables are split into resolved $\widetilde\phi$ and unresolved $\phi''$ components. Accordingly, any field $\phi$ may be expressed as follows:
\begin{equation}
\phi = \widetilde\phi + \phi''  
\label{LESFrame}
\end{equation}
It thus follows that the LES-filtered PIA system for dry air can be expressed as follows:
\begin{equation}
\frac{\partial \overline{\rho} \overline{\theta} \widetilde{u_j}}{\partial x_j} = 0
\label{Mass}
\end{equation}
\begin{equation}
\begin{aligned}
\frac{\partial \overline{\rho} \overline{\theta}\widetilde{U_i}}{\partial t} &= -\overline{\rho} \overline{\theta} C_p \widetilde{\theta}\frac{\partial \widetilde{\pi}}{\partial x_i}\\&-\underbrace{\frac{\partial \overline{\rho} \overline{\theta} \widetilde{U_i}\widetilde{U_j}}{\partial x_j}-\overline{\rho} g (\widetilde{\theta}-\overline{\theta})\delta_{i3} -2\overline{\rho} \overline{\theta} \epsilon_{ijk} \Omega_j(\widetilde{U_k}-U_{g,k})  -\frac{\partial \overline{\rho} \overline{\theta} \tau^{SGS}_{ij}}{\partial x_j}+{S_{m,i}}}_{M}
\end{aligned}
\label{Momentum}
\end{equation}

\begin{equation}
\begin{aligned}
\frac{\partial \overline{\rho} \overline{\theta} \widetilde{\theta}}{\partial t}+\frac{\partial \overline{\rho} \overline{\theta} \widetilde{\theta} \widetilde{U_j}}{\partial x_j} = -\frac{\partial \overline{\rho} \overline{\theta} \tau^{SGS}_{\theta}}{\partial x_j}
\end{aligned}
\label{Thermo}
\end{equation}
where $\Omega_j$ is the angular velocity vector of the Earth's rotation, $g$ is the gravitational acceleration, $U_{g,k}$ denotes the geostrophic wind velocity component,
$\tau^{SGS}_{ij}$ is the subgrid-scale (SGS) stress tensor, $\tau^{SGS}_{\theta}$ is the subgrid-scale heat flux, $\theta$ is the potential temperature, $\rho$ is the air density, $U_i$ is the air velocity component, $S_{m,i}$ is the momentum source term component such as the actuator model for wind turbine.
The solution of the coupled set of equations \eqref{Momentum}\eqref{Thermo} is facilitated by introducing an elliptic problem to ensure the anelastic constraint \eqref{Mass}:

\begin{equation}
\begin{aligned}
\frac{\partial M_j}{\partial x_j} = \frac{\partial \overline{\rho} \overline{\theta} \widetilde{\theta} Cp \frac{ \partial \widetilde{\pi}}{\partial x_j}}{\partial x_j} 
\end{aligned}
\label{Elliptic}
\end{equation}
The turbulence parameterisation in the model employs a mixing length closure scheme, which draws upon the seminal works of \citet{Redelsperger1981} and \citet{Cuxart2000}. This turbulence scheme is classified as a 1.5-order closure model, characterised by the incorporation of a prognostic equation for the turbulent kinetic energy (TKE) $k$. The turbulent subgrid scale terms are written as follows:
\begin{equation}
\begin{aligned}
\tau^{SGS}_{\theta}=-\frac{2}{3}\frac{L}{C_S}k^{\frac{1}{2}}\frac{\partial \widetilde{\theta}}{\partial x_i},
\end{aligned}
\label{SGStheta}
\end{equation}
\begin{equation}
\begin{aligned}
\tau^{SGS}_{ij}=\frac{2}{3}\delta_{ij}k-\frac{4}{15}\frac{L}{C_m}k^{\frac{1}{2}}\left(\frac{\partial \widetilde{U_i}}{\partial x_j}+\frac{\partial \widetilde{U_j}}{\partial x_i}-\frac{2}{3}\delta_{ij}\frac{\partial \widetilde{u_m}}{\partial x_m} \right)
\end{aligned}
\label{SGSM}
\end{equation}
with $C_m=4$ and $C_S=4$. The equation for the subgrid TKE is expressed as follows:
\begin{equation}
\begin{aligned}
\frac{\partial k}{\partial t}= -\frac{1}{\overline{\rho}}\frac{\partial \overline{\rho}  k \widetilde{U_j}}{\partial x_j} -\tau_{ij}^{SGS}\frac{\partial \widetilde{U_i}}{\partial x_j}+ \frac{g \tau_{\theta}^{SGS}}{\overline{\theta}}+\frac{1}{\overline{\rho}}\frac{\partial}{\partial x_j}\left(C_{2m}\overline{\rho} L k^{\frac{1}{2}}\frac{\partial k}{\partial x_j}\right )-C_{\epsilon}\frac{k^{\frac{3}{2}}}{L}
\end{aligned}
\label{TKE}
\end{equation}
with $C_{\epsilon}=0.85$ and $C_{2m}=0.2$.  The mixing length $L$ is the minimum mixing length between the horizontal grid cell $L_{\Delta}=(\Delta x\Delta y)^{1/2}$ and $L_{RM17}$ \citep{Rodier2017}:
\begin{equation}
\begin{aligned}
L=min(0.5L_{\Delta},L_{RM17})
\end{aligned}
\label{MixingL}
\end{equation}
The mixing length $L$ is designed to ensure that the turbulence scheme accurately balances subgrid and resolved turbulent exchanges across multiple scales. By allowing the turbulence scheme to adapt seamlessly across different resolutions and atmospheric conditions, this multi-scale approach ensures physical consistency and accuracy in the representation of turbulent processes, as demonstrated by \citet{Honnert2021}.

The equations are discretised using a staggered Arakawa C-grid to accurately represent spatial relationships. A fourth-order Runge-Kutta scheme is used for time integration to achieve good accuracy. Different advection schemes are used for momentum and scalar quantities. The momentum advection terms are calculated using a fourth-order scheme, which balances accuracy and efficiency. In contrast, scalar advection uses the piecewise parabolic method (PPM). This method is particularly effective at handling sharp gradients and maintaining distributions. For the elliptic problem, the horizontal components are addressed in Fourier space for efficiency, while the vertical components lead to a classical tridiagonal matrix \citep{Schumann1988}. 

The actuator method implemented in Meso-NH for modelling wind turbines has been successfully validated against the New Mexico wind tunnel experiments \citep{Joulin2020,Boumendil2024}. Furthermore, the model's capacity to represent the interactions between wind turbines in diverse stratified atmospheric boundary layer conditions has been corroborated through comparisons with in situ measurements from the SWiFT benchmark \citep{Jezequel2021}. This benchmark encompasses a range of atmospheric stability regimes, including stable, neutral, and unstable conditions, providing comprehensive data on inflow conditions, turbine response, and wake characteristics. The successful alignment of Meso-NH results with these measurements highlights its reliability in simulating wind turbine wakes under diverse atmospheric conditions.

\subsection{Setup Cases}

\begin{figure}
\centering
\def\svgwidth{0.45\textwidth}
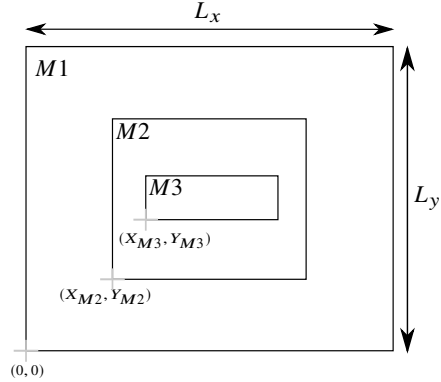
\caption{Three-domain nested simulation setup
with progressively refined spatial resolution (M1 → M2 → M3).}
\label{setup}
\end{figure}

The wake velocity deficit was evaluated for two wind turbine designs: a smaller turbine modelled after the NREL 5MW and a larger turbine based on the IEA 15MW model. To enable a direct comparison of the environmental impacts on each design, the hub height of the NREL 5MW turbine was adjusted to correspond to that of the IEA 15MW turbine. The key specifications for both turbines can be found in Table \ref{turb}. 
\begin{table}
\centering
\renewcommand{\arraystretch}{1.1} % Espacement vertical
\begin{tabular}{c@{\hspace{1cm}} c@{\hspace{1cm}} c@{\hspace{1cm}}}
Parameter & IEA 15MW & NREL 5MW \\
\noalign{\vskip 2mm}  % Ajoute 2mm d'espace vertical
Rotor diameter, $D$ (m) & 240 & 120\\
Hub height, $H_{hub}$ (m) & 150 & 150\\
Rated wind speed (m/s) & 10.59 & 11.2 \\
Maximum rotor speed (rpm) & 7.56 & 12\\
Minimal rotor speed (rpm) & 5 & 3.55 \\
\end{tabular}
\caption{Key specifications of the IEA 15MW and NREL 5MW}
\label{turb}
\end{table}
The simulations employ a three-level grid nesting strategy (Stein, 2000), with progressively refined meshes labelled M1 (coarsest), M2 (medium), and M3 (finest). The configuration of these three embedded domains is illustrated in Figure \ref{setup}. The specific simulation domain parameters for each turbine are detailed in Table \ref{tab:iea15mw} for the IEA 15MW and in Table \ref{tab:nrel5mw} for the NREL 5MW. This nested approach enables high-resolution simulation of small-scale features in the wake region ($60$ cells per diameter) while ensuring computational efficiency in the larger, less active areas of the flow field. The vertical resolution is consistent across all mesh levels for each turbine model. For the IEA 15MW turbine, a constant vertical resolution of $4$ m is maintained from the ground up to $350$ m. Similarly, for the NREL 5MW turbine, a $2$ m vertical resolution is used from the ground to $230$ m. This approach ensures detailed representation of the vertical structure in the lower atmosphere where the turbines operate. At altitudes above these heights, a vertical stretching technique is employed, exhibiting a 5$\%$ growth ratio. This results in a maximum cell size of $40$ m for the IEA 15MW and $20$ m for the NREL 5MW, extending to the domain top.

\begin{figure}
\centering
\begin{minipage}[t]{.48\linewidth}
\centering
\begin{tabular}{l|c|c|c}
\hline
\multicolumn{4}{c}{IEA 15 MW} \\
\hline
& \textbf{M1} & \textbf{M2} & \textbf{M3} \\
& \textbf{Coarse} & \textbf{Medium} & \textbf{Fine} \\
\hline
$L_x$ & 20 km & 10 km & 6 km \\
$L_y$ & 20 km & 10 km & 4 km \\
$L_z$ & 1.5 km & 1.5 km & 1.5 km \\
$\Delta x$ & 40 m & 8 m & 4 m \\
$\Delta y$ & 40 m & 8 m & 4 m\\
$\Delta z$ & 4 m$\rightarrow$40 m& 4 m$\rightarrow$40 m& 4 m$\rightarrow$40 m \\
$N_x$ & 500 & 1250 & 1500 \\
$N_y$ & 500 & 1250 & 1000 \\
$N_z$ & 148 & 148 & 148 \\
$X_M$   &  0  &  5 km  & 7 km  \\
$Y_M$   &  0  &  5 km  & 8 km \\
\hline
\end{tabular}
\captionof{table}{Specifications for IEA 15 MW turbine simulation domains}
\label{tab:iea15mw}
\end{minipage}%
\hfill
\begin{minipage}[t]{.48\linewidth}
\centering
\begin{tabular}{l|c|c|c}
\hline
\multicolumn{4}{c}{NREL 5 MW} \\
\hline
& \textbf{M1} & \textbf{M2} & \textbf{M3} \\
& \textbf{Coarse} & \textbf{Medium} & \textbf{Fine} \\
\hline
$L_x$ & 14.2 km & 7.2 km & 3 km \\
$L_y$ & 10 km & 5 km & 2 km \\
$L_z$ & 1.2 km & 1.2 km & 1.2 km \\
$\Delta x$ & 20 m & 10 m & 2 m \\
$\Delta y$ & 20 m & 10 m & 2 m \\
$\Delta z$ & 2 m$\rightarrow$20 m & 2 m$\rightarrow$20 m & 2 m$\rightarrow$20 m \\
$N_x$ & 720 & 720 & 1500 \\
$N_y$ & 500 & 500 & 1000 \\
$N_z$ & 192 & 192 & 192 \\
$X_M$   &  0  &  3.5 km  & 5.5 km  \\
$Y_M$   &  0  &  2.5 km  & 4 km \\
\hline
\end{tabular}
\captionof{table}{Specifications for NREL 5 MW turbine simulation domains}
\label{tab:nrel5mw}
\end{minipage}
\end{figure}

The simulations are carried out using the idealised framework of Meso-NH, where the flow is driven by geostrophic wind forcing and influenced by Coriolis forces. Horizontal periodic boundary conditions are imposed, while the top boundary features a rigid lid with a Rayleigh damping layer starting at $z=1.3$ km for the IEA 15 MW setup and at $z=1$ km for the NREL 5MW setup. At the bottom boundary, the velocity at the first grid point is determined based on Monin-Obukhov similarity theory, and a specified heat flux is applied to the potential temperature equation. The initial velocity field is uniform and matches the prescribed geostrophic wind. The initial potential temperature profile is set to 290 K from the surface up to 600 m, followed by a temperature inversion with a lapse rate of 20 K/km up to 800 m. Above this height, the free atmosphere is characterised by a stable stratification with a lapse rate of 6 K/km. Wind turbine power production is controlled by regulating the rotor's rotational speed and blade pitch using a simplified implementation of the ROSCO controller \citep{Abbas2022}, integrated into the Meso-NH framework. The wind turbine is represented using an Actuator Disk Rotation (ADR) model, which discretises the rotor disk into 60 radial elements and 200 azimuthal elements.

Three distinct atmospheric conditions — neutral, stable, and unstable — were simulated. The process started with a 20-hour simulation of a neutral atmospheric boundary layer using mesh M1. This simulation achieved convergence in the mean wind velocity at hub height after about 15 hours. The resulting data were then interpolated onto finer meshes, M2 and M3, for a subsequent 30-minute nested simulation without a wind turbine, followed by a 30-minute nested simulation with a wind turbine. A single wind turbine was positioned at specific coordinates depending on the turbine model: at (9.5 km, 10 km) for the IEA 15MW case and at (6.7 km, 5 km) for the NREL 5MW case. To simulate stable and unstable atmospheric conditions, surface heat fluxes of $-9 W/m^2$ and $90 W/m^2$ were applied, respectively, to the initially converged neutral simulation on mesh M1. These fluxes were maintained over a 5-hour period. From this initial simulation, a one-hour interval with minimal fluctuations in velocity (approximately $0.5$ m/s) and wind direction (around $0.5\degree$) was selected. This specific time period was then re-simulated with the inclusion of additional meshes (M2 and M3). The nested simulation was conducted in two phases: the first 30 minutes without a wind turbine, followed by another 30 minutes with a wind turbine present, consistent with the methodology used for the neutral case. All simulations were conducted with the following shared parameters: a surface roughness of $z_0 = 50$ mm, geographical coordinates of $33.3\degree$ latitude and $-119.5\degree$ longitude, and geostrophic wind components $(U_G, V_G) = (11.04, -4.81)$ m/s. 

\subsection{Results}
\subsubsection{Statistical inflow characterization and turbine performance indicators}

The analysis focuses on the final 10 minutes of each simulation. Upstream flow characteristics are evaluated by computing the mean velocity and turbulence intensity at a location 4$D$ upstream of the turbine, within a disk of 1$D$ diameter centered at hub height. The statistical properties of the upstream flow are summarized in Table \ref{tab:upstream_stats} for the IEA 15MW and Table \ref{tab:upstream_stats_2} for the NREL 5MW, with results shown for 10-minute window. The thrust coefficients fall within the expected range reported in the literature \citep{Gaertner2020,Abbas2022}. 
\begin{table}
 \centering
 \renewcommand{\arraystretch}{1.2}
 \begin{tabular}{l|c|c|c}
 \hline
 &{\textbf{Stable}} &{\textbf{Neutral}} & {\textbf{Unstable}} \\
 \hline
 $ I_u $ [\%]         & 3.0    & 7.0   & 6.8    \\
 $ I_v $ [\%]         & 1.5    & 6.3   & 7.0   \\
 $ I_w $ [\%]         & 0.8    & 5.6   & 7.5    \\
 $ A^E_v $ [s]        & 20.0   & 5.0   & 11.0   \\
 $ A^E_w $ [s]        & 6.0    & 3.4   & 9.0   \\
 $ TI $ [\%]          & 2.0    & 6.3   & 7.0  \\
 $ U_{\infty} $ [m/s] & 11.2   & 10.2  & 9.65  \\
 $ C_T$               & 0.78   & 0.73  & 0.79   \\
 \hline
 \end{tabular}
 \caption{Statistical characteristics of upstream flows for each simulated atmospheric condition of the IEA 15MW setup. Values are given for 10 min window.}
 \label{tab:upstream_stats}
 \end{table}
\begin{table}
\centering
\renewcommand{\arraystretch}{1.2}
\begin{tabular}{l|c|c|c}
\hline
& {\textbf{Stable}} & {\textbf{Neutral}} & {\textbf{Unstable}} \\
\hline
$ I_u $ [\%]         & 6.2     & 8.0    & 6.5    \\
$ I_v $ [\%]         & 5.5     & 7.1    & 6.9    \\
$ I_w $ [\%]         & 4.6     & 6.6    & 6.7    \\
$ A^E_v $ [s]        & 2.2     & 4.     & 27.    \\
$ A^E_w $ [s]        & 1.7     & 3.     & 3.9    \\
$ TI $ [\%]          & 5.4     & 7.2    & 6.7    \\
$ U_{\infty} $ [m/s] & 10.2    & 10.    & 9.7    \\
$ C_T$               & 0.84    & 0.71   & 0.83   \\
\hline
\end{tabular}
\caption{Statistical characteristics of upstream flows for each simulated atmospheric condition of the NREL 5MW setup. Values are given for 10 min window.}
\label{tab:upstream_stats_2}
\end{table}
For all NREL 5MW cases, the mean velocity is approximately 10~m/s, whereas a higher mean velocity is observed for the IEA 15MW under stable conditions. This difference stems from the spatial averaging method: under stable atmospheric conditions, the velocity profile exhibits a steeper gradient near the surface. The smaller rotor diameter of the NREL 5MW limits the averaging to regions with lower velocities, while the larger rotor of the IEA 15MW encompasses more of the higher-velocity region near the surface, resulting in an increased mean velocity. As expected, stable cases display lower overall turbulence intensity. In contrast, the unstable case shows a turbulence intensity similar to the neutral case, but with a reduced streamwise component due to enhanced vertical mixing, which diminishes streamwise velocity fluctuations. This redistribution of turbulence intensity with atmospheric stability is consistent with previous findings \citep{Du2021}.

The integral time scales are significantly greater for the IEA 15MW, as expected given the averaging over a broader rotor area. This is due to the IEA 15MW rotor spanning a wider vertical range (30 m to 270 m), thus including lower altitudes characterized by larger integral time and length scales, whereas the NREL 5MW rotor (90 m to 210 m) excludes these near-surface layers. Consequently, the spatial averaging for the IEA 15MW is weighted toward the higher integral scales near the ground, resulting in increased overall integral time scales relative to the NREL 5MW. In a broader context, the effect of different averaging periods on upstream flow statistics has been investigated. The minimal variation observed between the 10-minute and 20-minute averaging windows indicates that the flow statistics remain stable over the entire analysis period.

\subsubsection{Neutral cases}

\begin{figure}
    \centering
    \begin{subfigure}{0.9\textwidth}
        \centering
        \includegraphics[width=\linewidth]{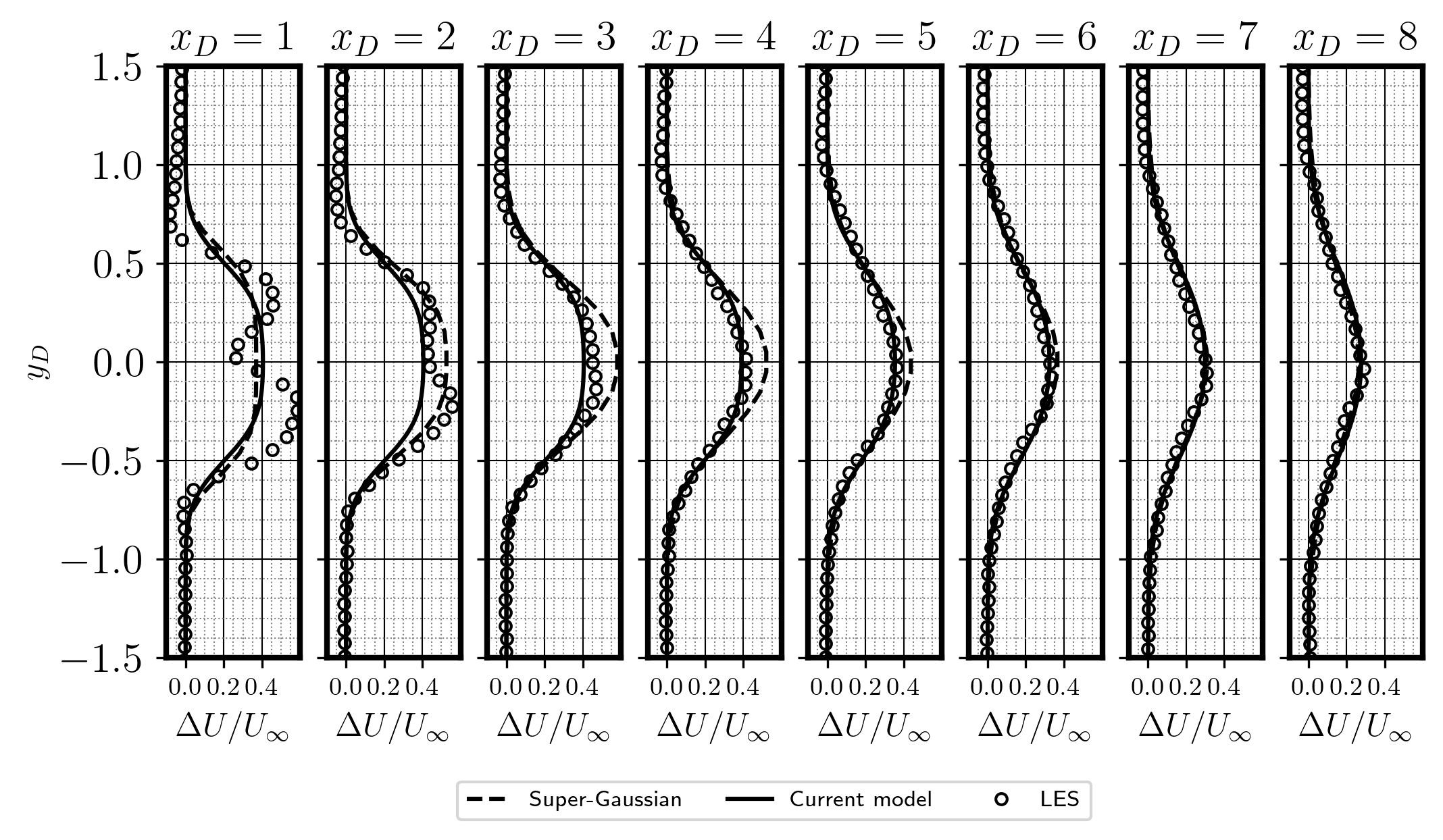}
        \caption{IEA 15MW Neutral case - 10 minutes window}
        \label{fig:ADR1_10}
    \end{subfigure}
    \vspace{1em}
    \begin{subfigure}{0.9\textwidth}
        \centering
        \includegraphics[width=\linewidth]{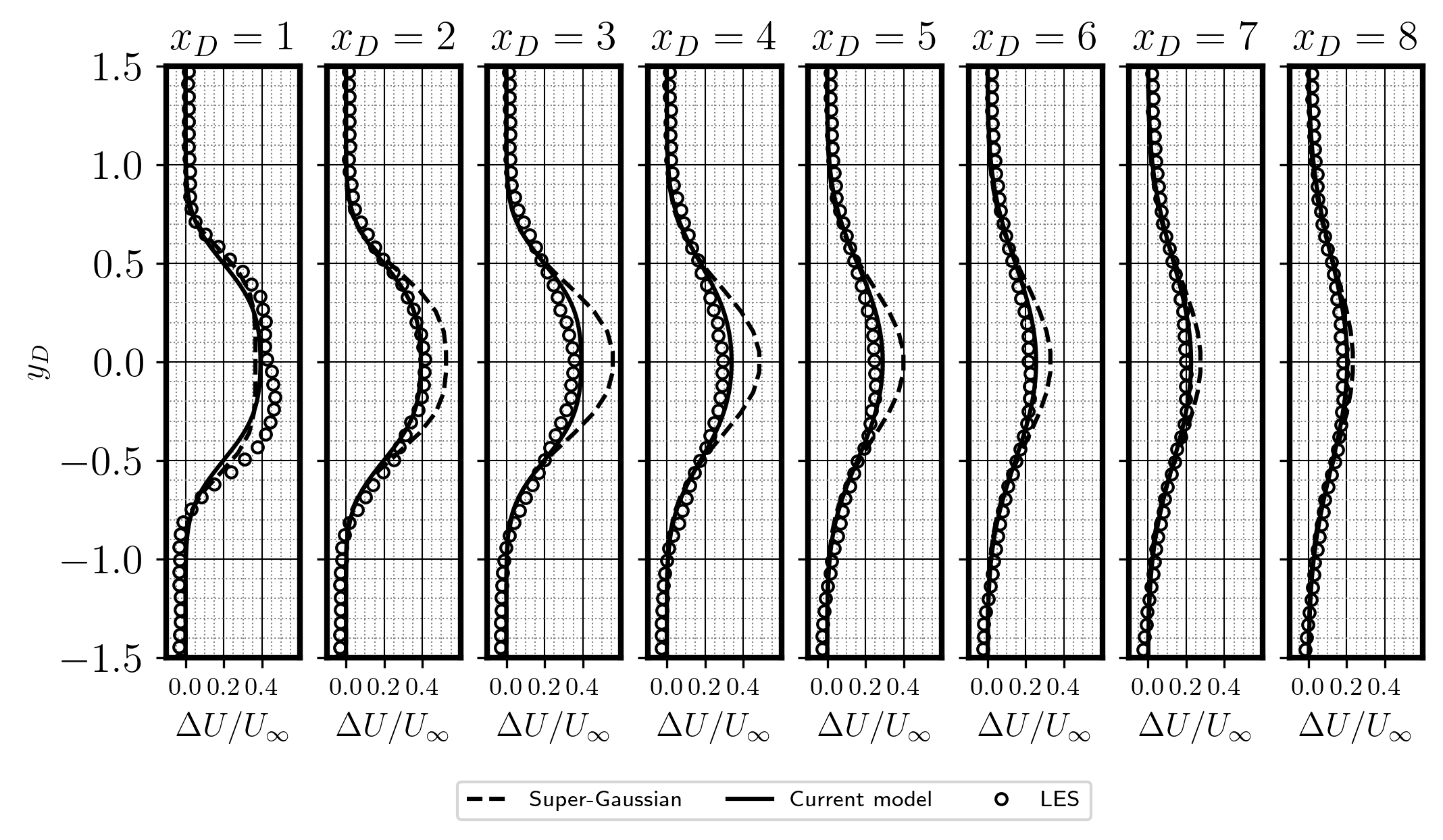}
        \caption{NREL 5MW Neutral case - 10 minutes window}
        \label{fig:ADR4_10}
    \end{subfigure}
    \vspace{1em}
    \caption{Assessment of velocity deficit prediction accuracy in neutral stability scenarios}
    \label{fig:Neutral}
\end{figure}

LES results are compared with the estimates from both the Super-Gaussian model and the Current model. The super-Gaussian model described in \citealp{Blondel2020} is applied using input parameters from the inflow statistics ($I_u,U_{\infty}$) and the thrust coefficient ($C_T$), as provided in Tables \ref{tab:upstream_stats} and \ref{tab:upstream_stats_2}. Figure \ref{fig:Neutral} shows the downstream velocity deficit for the IEA 15MW and NREL 5MW turbines under neutral atmospheric conditions, averaged over a 10-minute period. Close to the turbine, the wake displays a double-peaked (double-Gaussian) structure, reflecting the influence of individual blade wakes before they merge further downstream—a feature not captured by either model. Further downstream, for both turbines, the Super-Gaussian model tends to underestimate wake spreading, resulting in a slightly higher velocity deficit at the wake center compared to LES. This effect is more noticeable for the NREL 5MW turbine, as illustrated in Figure \ref{fig:ADR4_10}: a modest increase in streamwise turbulence intensity—from 7\% (IEA 15MW) to 8\% (NREL 5MW)—leads to enhanced wake spreading in the LES results, which the Super-Gaussian model does not fully reflect. While the Super-Gaussian model offers reliable predictions that are largely insensitive to small fluctuations in turbulence, this characteristic can limit its ability to capture subtle but meaningful variations in wake behaviour observed in LES. In contrast, the current model incorporates inflow turbulence parameters through Taylor’s diffusion theory, allowing it to respond to these minor variations. This sensitivity underscores the importance of accounting for inflow conditions to achieve accurate wake predictions and a realistic representation of atmospheric variability.

\begin{figure}
    \centering
    \includegraphics[width=\linewidth]{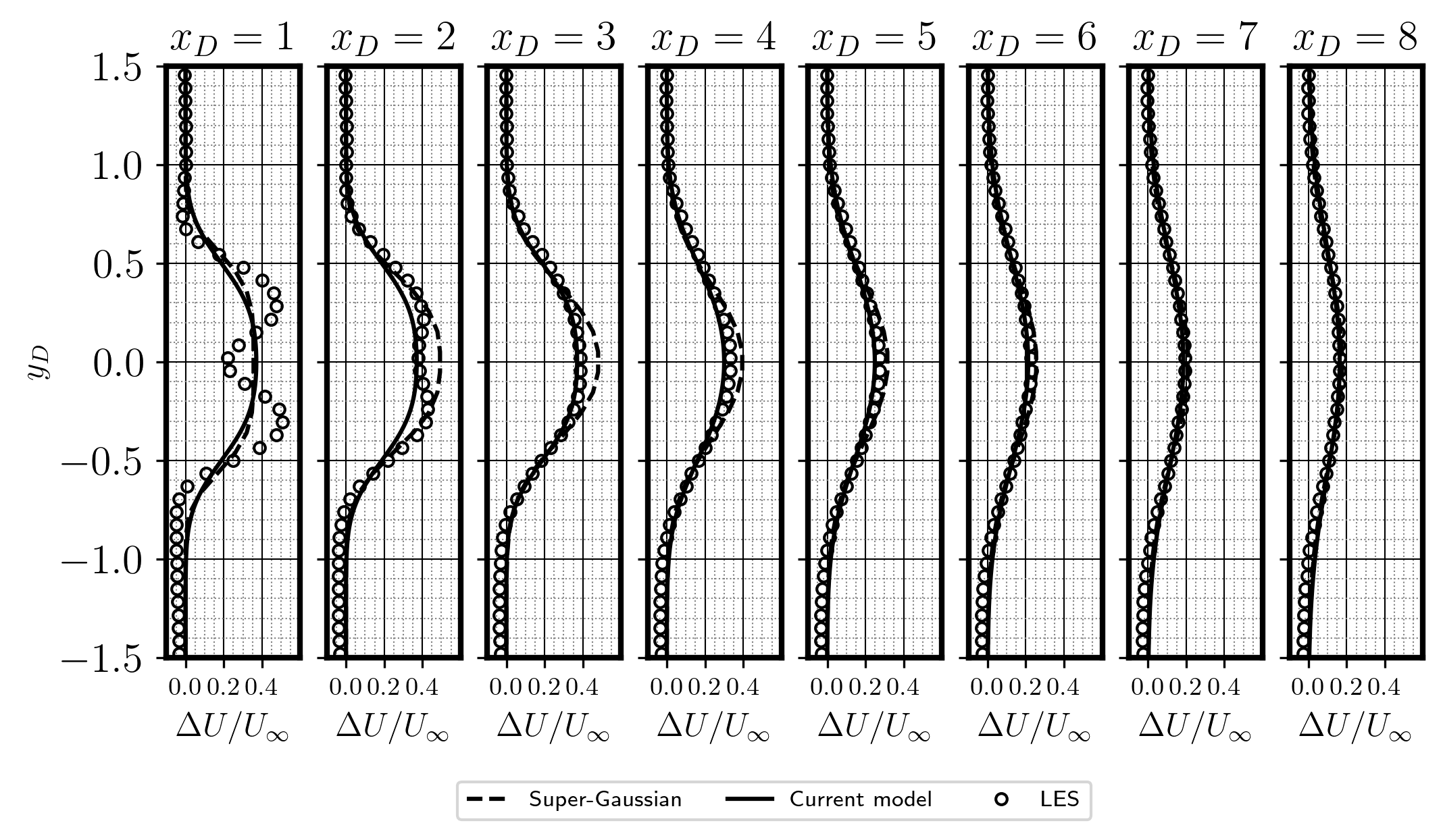}
    \caption{Velocity deficit predictions under neutral stability from LES data of \citet{Jezequel2023}.}
    \label{fig:Neutral1}
\end{figure}

\begin{figure}
    \centering
    \includegraphics[width=\linewidth]{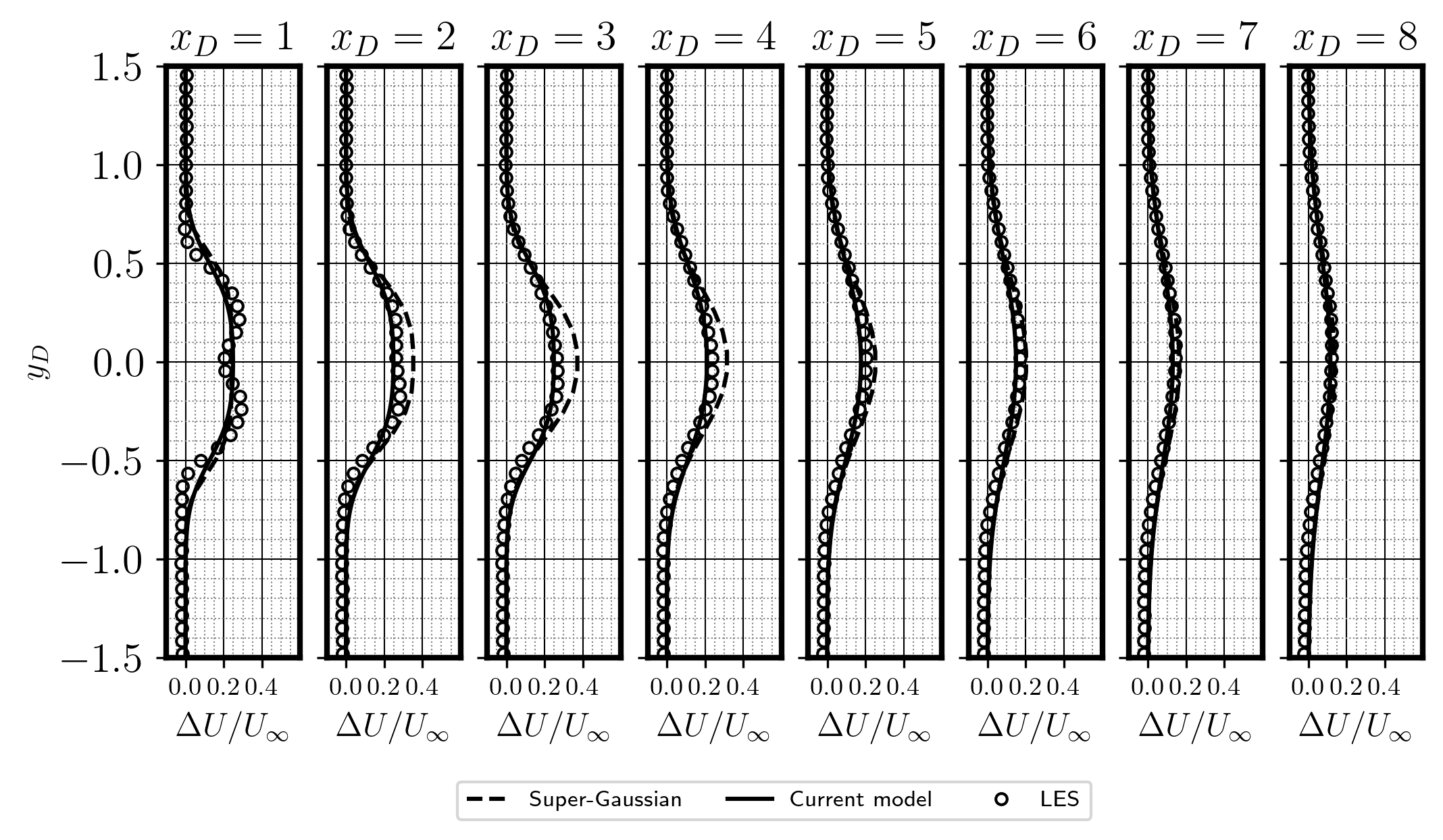}
    \caption{Velocity deficit predictions under neutral stability with blade pitching ($C_T=0.5$) from LES data of \citet{Jezequel2023}.}
    \label{fig:Pitch4}
\end{figure}

\begin{figure}
    \centering
    \includegraphics[width=\linewidth]{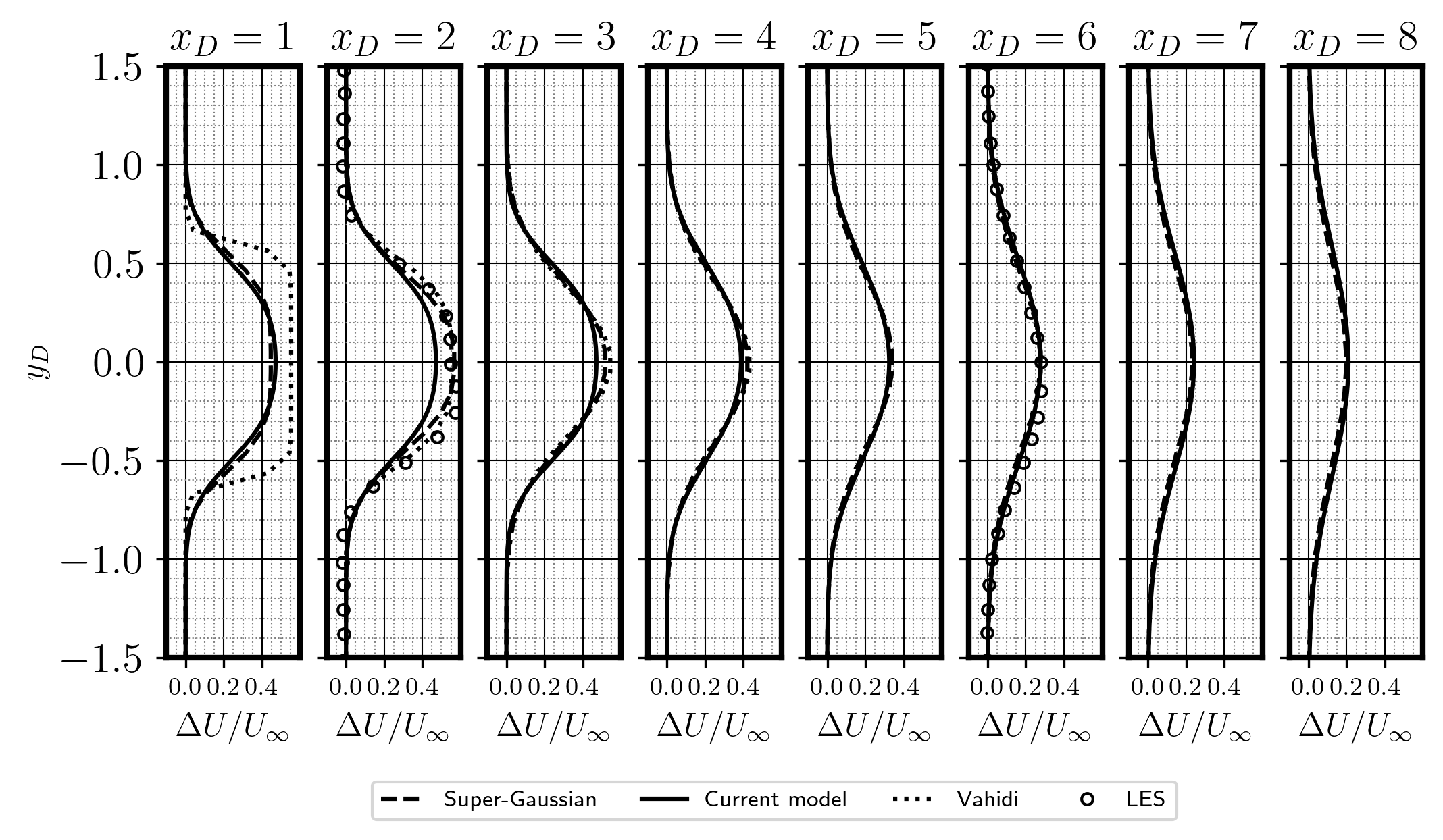}
    \caption{Velocity deficit predictions under neutral stability from LES data of Case2 of \citet{Vahidi2022}.}
    \label{fig:Case2}
\end{figure}

Figure \ref{fig:Neutral1} presents the downstream velocity deficit obtained from LES simulations conducted within the MOMENTA project \citep{Jezequel2023}. The details of the simulation setup are described in \citet{Jezequel2024b}. For this case, the streamwise turbulence intensity is 11.2\%. It is noteworthy that the Super-Gaussian model performs quite well under these conditions. The current model shows an even closer match, exhibiting good agreement with the LES results. Figure \ref{fig:Pitch4} presents results for the blade pitching scenario, used to evaluate the current model’s ability to represent atypical thrust coefficients ($C_T = 0.5$). To further assess the model’s robustness in cases where some input data are unavailable, the Case 2 from \citet{Vahidi2022} is also included. In this case, the integral time scales are not provided in the original paper, therefore they are here estimated based on the previously simulated neutral atmospheric conditions. The estimated values for both the lateral and vertical integral time scales are approximately 5 seconds ($A^E_v = A^E_w = 5$). Figure \ref{fig:Case2} compares the velocity deficit from LES, the Vahidi model, the Super-Gaussian model, and the Current model. The results are generally consistent across all models. However, a discrepancy for the current model is observed in the near wake region ($x_D = 2$), where it underestimates the velocity deficit. This deviation can likely be attributed to uncertainties in the estimated integral time scales as well as in the thrust coefficient. Overall, the model performs quite well demonstrating its robustness.

\subsubsection{Stable cases}

\begin{figure}
    \centering
    \begin{subfigure}{0.9\textwidth}
        \centering
        \includegraphics[width=\linewidth]{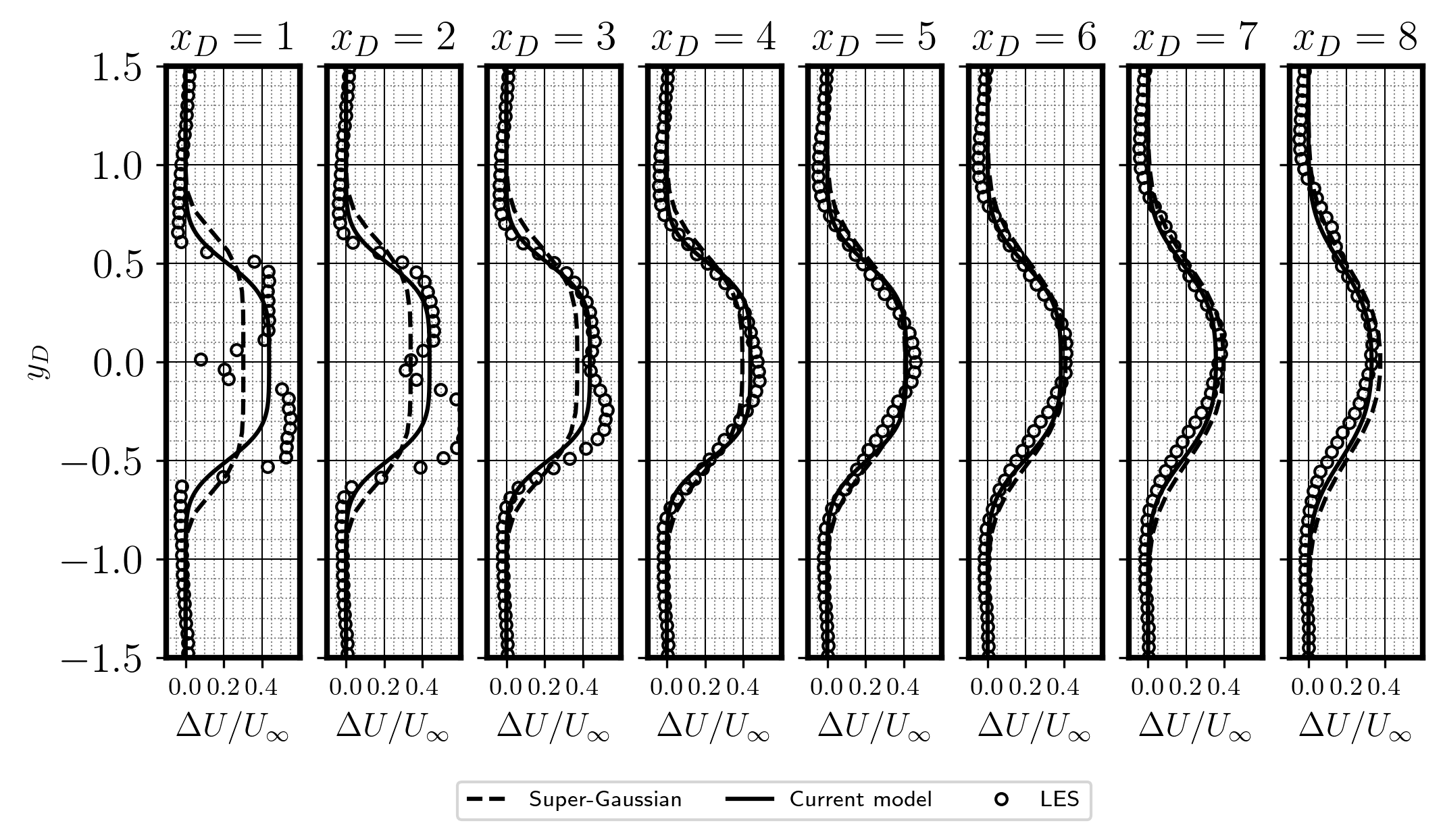}
        \caption{IEA 15MW Stable case - 10 minutes window}
        \label{fig:ADR2_10}
    \end{subfigure}
    \vspace{1em}
    \begin{subfigure}{0.9\textwidth}
        \centering
        \includegraphics[width=\linewidth]{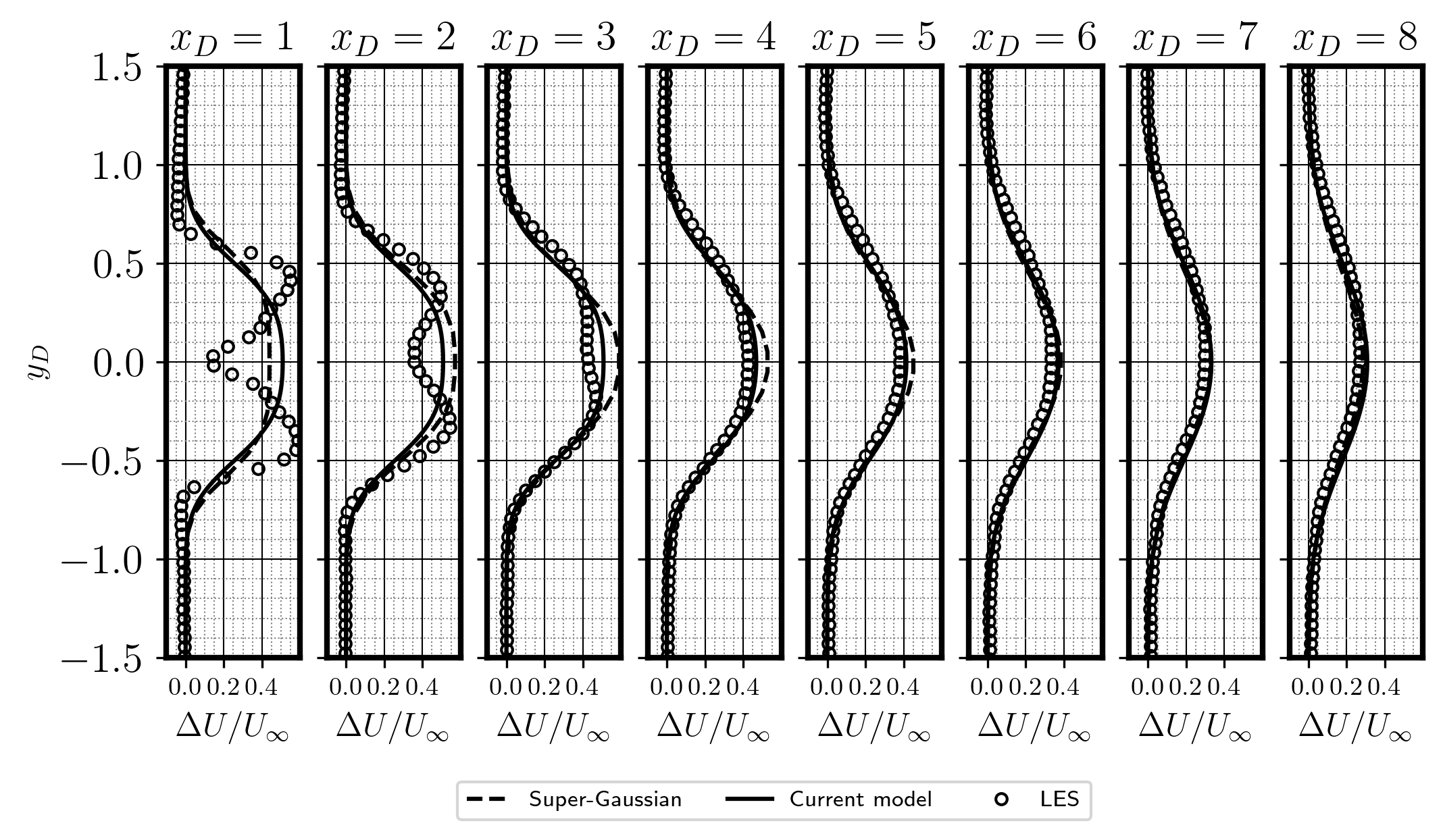}
        \caption{NREL 5MW Stable case - 10 minutes window}
        \label{fig:ADR6_10}
    \end{subfigure}
    \vspace{1em}
    \caption{Assessment of velocity deficit prediction accuracy in stable stability scenarios}
    \label{fig:Stable}
\end{figure}

LES results are compared with predictions from the Super-Gaussian model and the Current model. Figure \ref{fig:Stable} shows the downstream velocity deficit for both the IEA 15MW and NREL 5MW turbines under stable atmospheric conditions, averaged over 10 minutes. In the IEA 15MW case (Figure \ref{fig:ADR2_10}), the wake exhibits asymmetric behaviour, with the maximum velocity deficit shifted away from the centerline. This shift is attributed to a small angle (approximately $1^\circ$) between the wind direction and the streamwise axis —an issue that is challenging to control within the Meso-NH framework. This slight misalignment leads to both the observed offset and changes in the size of the initial double peak in the wake profile. In contrast, this effect is not present in Figure \ref{fig:ADR6_10}, where the wind alignment is nearly perfect.
%Overall, the current model agrees well with the LES results, as does the Super-Gaussian model. For the NREL 5MW case, the Super-Gaussian model slightly overestimates the velocity deficit, but the overall agreement remains strong. This is somewhat unexpected, as the Super-Gaussian model is not specifically designed to account for varying atmospheric stability. However, in the IEA case, the low streamwise turbulence intensity provides sufficient information for the Super-Gaussian model to perform accurately. For the NREL case, the streamwise turbulence intensity is similar to that of a neutral case, though the integral time scales are lower. Additionally, the lateral and vertical turbulence intensities remain relatively high, making the turbulence structure comparable to neutral conditions. This likely explains why the Super-Gaussian model performs relatively well in this scenario. 
Overall, the current model shows good agreement with the LES results, similarly to the Super-Gaussian model. For the NREL 5MW case, the Super-Gaussian model slightly overestimates the velocity deficit, but the discrepancy remains minor. This outcome is somewhat unexpected given that the Super-Gaussian model is not explicitly designed to accommodate varying atmospheric stability regimes. Regarding the NREL case, the turbulence statistics appear to be analogous to those of the neutral atmospheric conditions for which the Super-Gaussian model was originally calibrated, thereby explaining its satisfactory performance in this scenario. In contrast, for the IEA case, it is the very low streamwise turbulence intensity that provides a well-defined constraint, enabling the Super-Gaussian model to accurately capture the wake deficit decay.

These results demonstrate the robustness of the Super-Gaussian model across both neutral and stable conditions. The current model, by explicitly incorporating inflow turbulence characteristics, further enhances predictive accuracy and shows greater sensitivity to subtle variations in atmospheric conditions, making it a valuable tool for capturing a wider range of wake behaviours.

\subsubsection{Unstable cases}

\begin{figure}
    \centering
    \begin{subfigure}{0.9\textwidth}
        \centering
        \includegraphics[width=\linewidth]{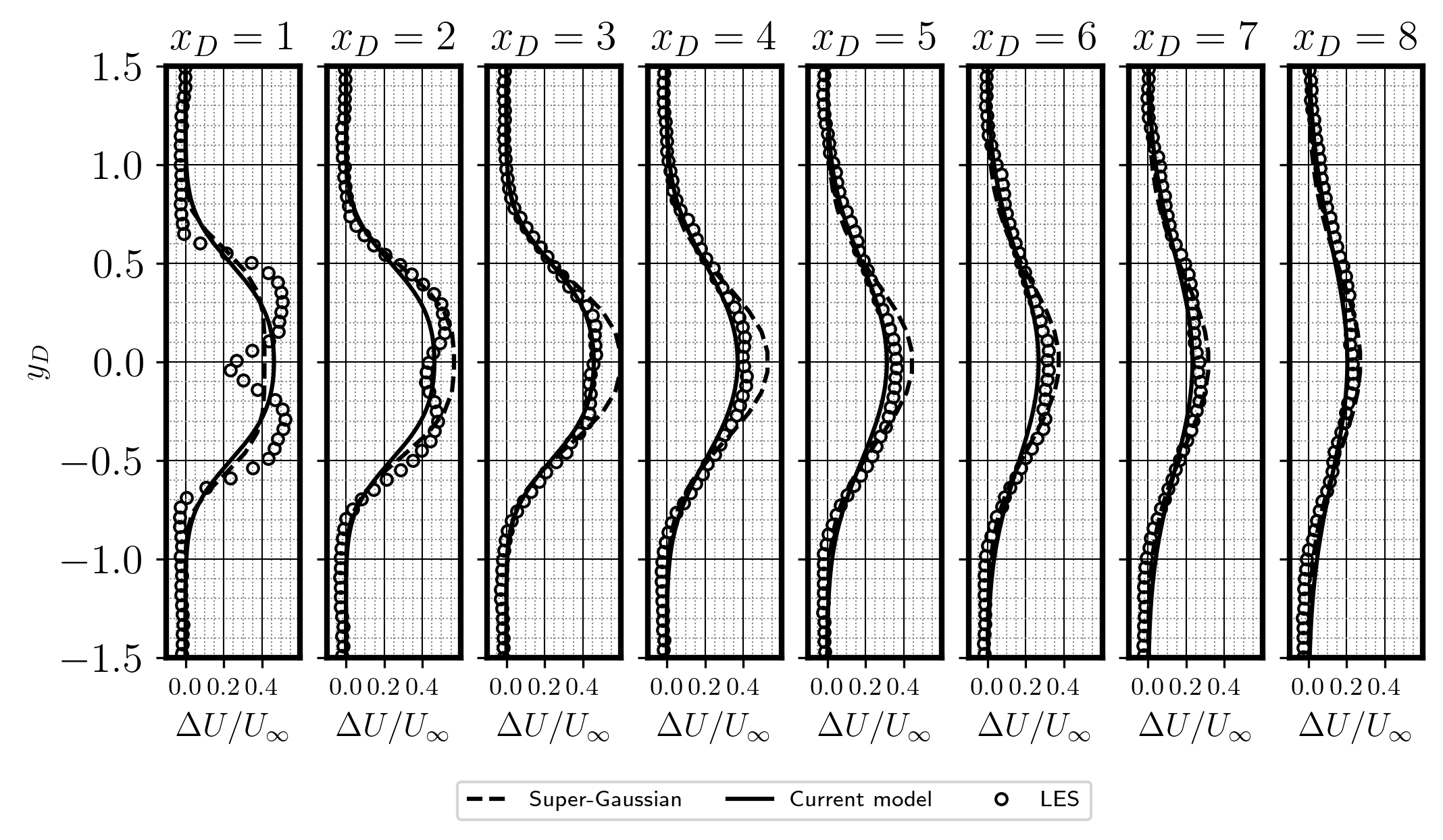}
        \caption{IEA 15MW Unstable case - 10 minutes window}
        \label{fig:ADR3_10}
    \end{subfigure}
    \vspace{1em}
    \begin{subfigure}{0.9\textwidth}
        \centering
        \includegraphics[width=\linewidth]{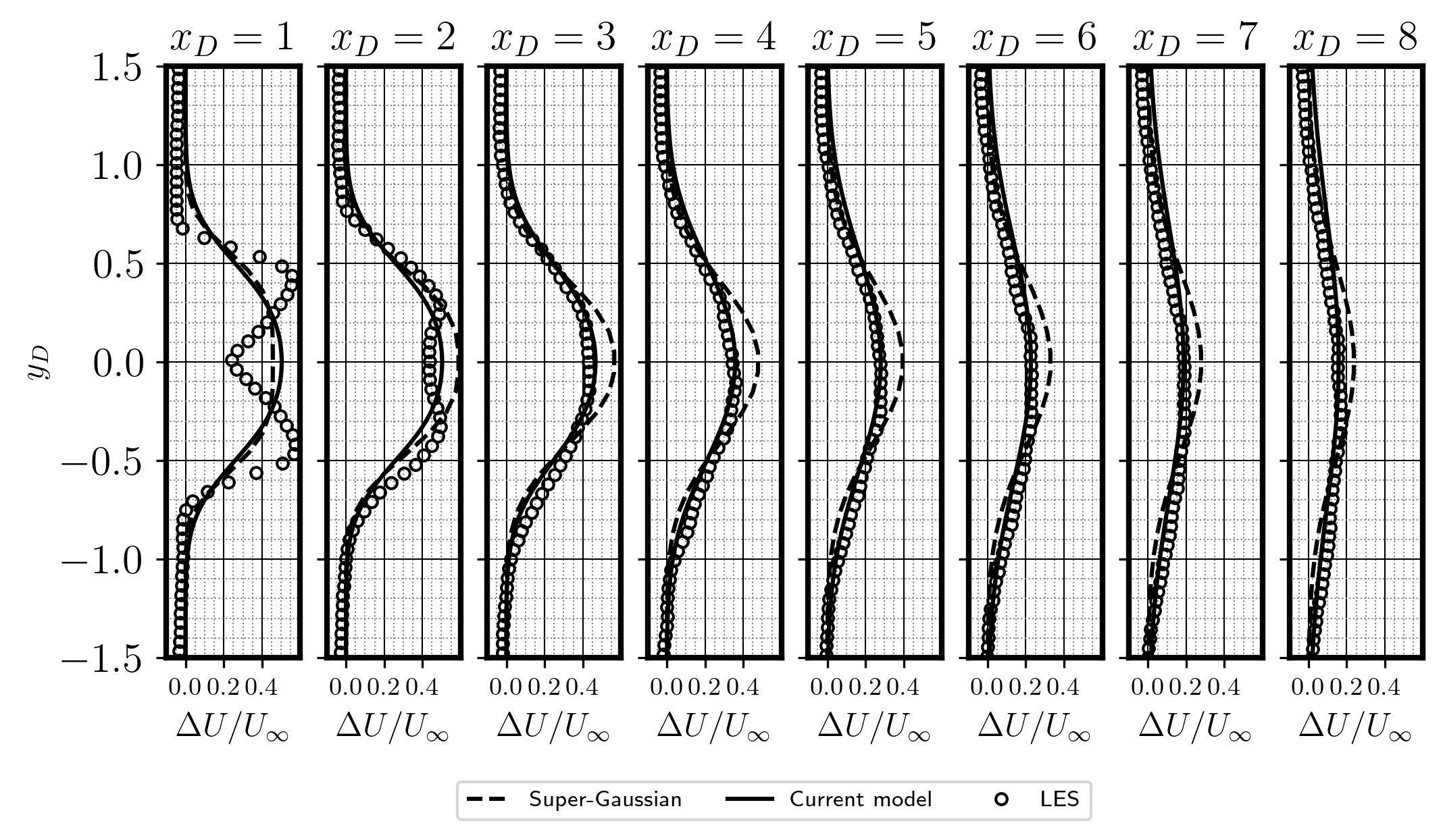}
        \caption{NREL 5MW Unstable case - 10 minutes window}
        \label{fig:ADR5_10}
    \end{subfigure}
    \vspace{1em}
    \caption{Assessment of velocity deficit prediction accuracy in unstable stability scenarios}
    \label{fig:Unstable}
\end{figure}

LES results are compared with the estimates from both the Super-Gaussian model and the current model. Figure \ref{fig:Unstable} presents the downstream velocity deficit for both the IEA 15MW and NREL 5MW turbines under unstable atmospheric conditions, averaged over 10 minutes. The Super-Gaussian model tends to overestimate the velocity deficit, as previously observed for the neutral case (Figure \ref{fig:ADR4_10}). However, for the IEA 15MW turbine, it captures the correct behaviour in the far wake region ($x_D > 6$). The performance is somewhat less accurate for the NREL 5MW case, where the Super-Gaussian model consistently overestimates the wake deficit at all downstream distances, including the far wake.

By comparison, the current model shows excellent agreement with the LES results. A key feature of unstable atmospheric conditions is the presence of higher integral time scales in the lateral and vertical directions, as well as greater lateral turbulence intensity compared to the streamwise component. This turbulent structure accelerates wake recovery, an effect that the Super-Gaussian model does not adequately capture. In contrast, the current model effectively represents this enhanced wake recovery, as it is grounded in the physical processes of turbulent diffusion.

\begin{figure}
    \centering
    \includegraphics[width=\linewidth]{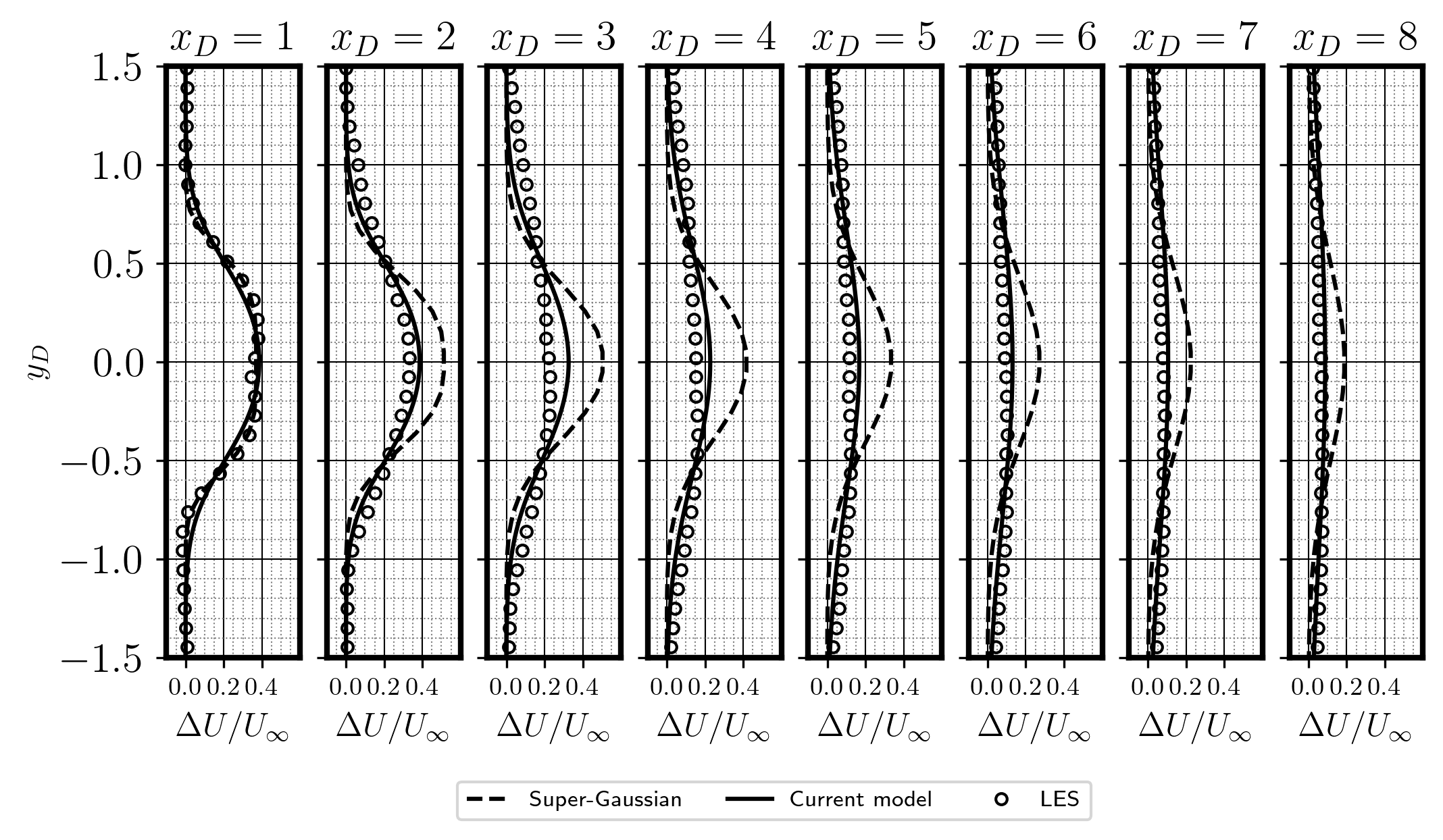}
    \caption{Velocity deficit predictions under unstable (named "Unstable") stability from LES data of \citet{Jezequel2023}.}
    \label{fig:Unstable2}
\end{figure}

Figure \ref{fig:Unstable2} shows the downstream velocity deficit from LES simulations performed as part of the MOMENTA project \citep{Jezequel2023} for an unstable atmospheric scenario ("Unstable" case). The flow is highly turbulent, with intensities of $I_u=10\%$, $I_v=16\%$, and $I_w=8.7\%$. LES results are compared with predictions from both the Super-Gaussian model and the current model. Despite the high streamwise turbulence intensity, the Super-Gaussian model does not capture the rapid wake recovery observed in the LES. As previously discussed, this limitation is related to the elevated lateral and vertical turbulence intensities, which enhance wake mixing and accelerate recovery. The current model is specifically designed to account for anisotropic turbulent diffusion processes, making it well-suited to reproduce this type of behaviour.

\section{Discussion and Conclusion}\label{Conclusion}

The wake deficit model presented here is based on the principle that a wind turbine wake initially features a uniform velocity deficit due to the rotor’s low-pass filtering effect, which then spreads through turbulent mixing. Turbulence is treated as a diffusion process, following the classical framework of \citet{Taylor1922}, which physically describes fluid parcel dispersion. In addition to turbulent diffusion, the model accounts for turbine-induced mixing effects by representing them as the growth of a mixing layer, following \citet{Vahidi2022}. Mathematically, the model represents the turbulent dispersion of the initial velocity deficit as the convolution of two probability density functions. This convolution produces an analytical solution for the normalised wake velocity profile, $f(r/R(x))$, notably without assuming a fixed shape for the deficit—an improvement over traditional empirical models. 

Determining the actual velocity deficit requires estimating the normalisation factor $U_w(x)$ (or more generally, $\alpha_x$). This is achieved by applying signal theory, which explains how turbulent diffusion—modelled as a Gaussian probability density function—attenuates the initial wake deficit predicted by one-dimensional momentum theory. This approach enables the analytical determination of the normalisation factor, ensuring the model remains fully analytical and physically grounded, while still being straightforward to implement computationally. A key practical aspect in implementing the model is estimating the wake travel time. Because travel time depends on the velocity deficit, which itself evolves with travel time, the problem is weakly coupled. Nevertheless, this coupling is sufficiently mild to permit a simplified iterative procedure: both travel time and velocity deficit are updated twice, beginning from the convective velocity estimated via momentum theory. This approach balances accuracy and computational efficiency. The implementation steps are detailed in Appendix \ref{Tech}.

The model has been validated across a range of atmospheric stability regimes—including stable, neutral, and unstable conditions—using LES data from this study and from the literature, covering various turbine sizes. Results show good agreement across all cases. The model’s accuracy surpasses that of advanced empirical approaches such as the Super-Gaussian model, although the latter remains robust and effective when only streamwise turbulence intensity is available. A key strength of the present model is its explicit mathematical formalisation of the physical processes governing wake evolution, accounting not only for velocity and streamwise turbulence intensity, but also for the detailed structure of the inflow. This comprehensive framework enables highly accurate predictions of wake velocity deficits. However, a significant limitation is the model’s reliance on detailed inflow information. As a result, the main challenge in wake modeling shifts toward the accurate characterization of inflow conditions.

Currently, on-site wind measurements for wind energy are still dominated by meteorological masts equipped with cup anemometers. These instruments provide reliable mean wind speed and axial turbulence intensity at specific heights but do not capture the full three-dimensional turbulence structure, particularly lateral and vertical fluctuations \citep{IEC61400-50-1}. Sonic anemometers, while capable of measuring all three wind components, are less robust and more costly, which limits their widespread use. Doppler LIDAR systems \citep{IEC61400-50-4} offer multi-height wind profiling, but significant uncertainties remain in reconstructing lateral and vertical turbulence components, with accuracy highly dependent on retrieval algorithms. No current measurement technology provides robust, site-specific data for all turbulence components and their time scales \citep{Shaw2022,Kosovic2025}. Furthermore, reanalysis products such as ERA5 do not resolve the 3D turbulence structure, limiting their utility for primary yield assessment.

This context underscores the need for further research to improve measurement techniques or enhance the available wind data, as advances in wake modeling are fundamentally dependent on the quality of inflow characterisation. Ultimately, the predictive capability of any physical model depends not only on a rigorous representation of the governing mechanisms but also on the precision and completeness of the environmental input parameters. Without high-quality inflow measurements, even the most advanced models are fundamentally constrained. Nevertheless, the analytical framework developed here shows that, when detailed inflow information is available, it can deliver significantly improved wake deficit predictions compared to empirical approaches. In cases where only limited turbulence data are accessible, the Super-Gaussian model remains a reliable and pragmatic choice. Ultimately, the choice of modelling approach should reflect the quality and completeness of the available input data, highlighting the value of continued progress in both measurement techniques and physical modelling.

%In situations where information is limited, the Super-Gaussian framework has demonstrated its robustness and ability to represent wake deficits as accurately as possible with the available data.

\section*{Acknowledgements}

This project was provided with computer and storage resources by GENCI at 
TGCC thanks to the grant 2024-A0150114592 on the supercomputer 
Joliot Curie's the SKL partition. We would like to thank our colleague Dr. F. Blondel for valuable discussions on these results, and in particular for his input on the comparison with the Super-Gaussian formulation.

\section*{Funding}
This work was supported by internal funding from IFP Energies nouvelles (IFPEN) dedicated to the Wind Energy Program.

\section*{Declaration of interests}
The authors report no conflict of interest.

\section*{Author ORCIDs}
Emeline No\"el: \url{https://orcid.org/0000-0003-2429-7737} \\
Erwan J\'ez\'equel: \url{https://orcid.org/0000-0001-5024-364X}\\
Pierre-Antoine Joulin: \url{https://orcid.org/0000-0002-3398-3395}

\section*{Author Contributions}

E. No\"el designed the study and wrote the manuscript.  
E. J\'ez\'equel and P.-A. Joulin contributed to the review and editing of the manuscript.

\appendix
\section{Technical Steps for Model Implementation} \label{Tech}

In this section, the algorithm employed to determine the wake velocity deficit at a downstream position ($x$) using the current wake model is described.
\\

\begin{enumerate}
    \item \textbf{Gather Input Parameters:}
    \begin{itemize}
        \item \( \sigma_v, \, \sigma_w \): Standard deviations of lateral and vertical turbulent velocities
        \item \( A_v^E, \, A_w^E \): Eulerian integral length scales (lateral and vertical)
        \item \( U_{\infty} \): Free-stream velocity
        \item \( C_T \): Thrust coefficient
        \item \( D \): Rotor diameter
    \end{itemize}

    \item \textbf{Compute the Lagrangian Integral Time Scales:}
    \begin{align}
        A_v^L &= A_v^E\, U_{\infty} \frac{\gamma}{\sigma_v}, \\
        A_w^L &= A_w^E\, U_{\infty} \frac{\gamma}{\sigma_w},
    \end{align}
    where the stability-dependent coefficient \( \gamma \) is defined as
    \begin{equation}
    \gamma = 
    \begin{cases}
    0.4 & \text{neutral/weakly stable conditions} \\
    0.6 & \text{convective conditions}.
    \end{cases}
    \end{equation}

    \item \textbf{Initial Path Length at Reference Location \( x_0 = 1D \):}
    
    \begin{enumerate}
        \item The minimal characteristic time scale is:
        \begin{equation}
            T_0 = \frac{1}{\frac{1}{2}(1 + \sqrt{1 - C_T})}.
        \end{equation}

        \item Define the path length function:
        \begin{equation}
        \mathcal{P}(A^L, \sigma, T) = \frac{\sigma}{D} \left( \sqrt{2A^L T - 2(A^L)^2\left[1 - e^{-T/A^L}\right]} + 2S(U_{\infty} T - (x - x_0)) \right),
        \end{equation}
        with spreading parameter \( S = 0.043 \).

        \item The initial normalised path lengths are:
        \begin{align}
            \sigma_{{all}_v}^0 &= \mathcal{P}(A_v^L, \sigma_v, T_0), \\
            \sigma_{{all}_w}^0 &= \mathcal{P}(A_w^L, \sigma_w, T_0), \\
            \sigma_{all}^0 &= \sqrt{\sigma_{{all}_v}^0\, \sigma_{{all}_w}^0}.
        \end{align}
    \end{enumerate}

    \item \textbf{Iterative Estimation of Wake Deficit at Downstream Location \( x > x_0 \):}
    
    The wake velocity and total deficit are computed via an iterative process:

    \begin{enumerate}
        \item \textit{Initial guess for characteristic time:}
        \begin{equation}
            T^{(0)} = \frac{x - x_0}{\frac{1}{2}(1 + \sqrt{1 - C_T})}.
        \end{equation}

        \item \textit{Initial guess for path lengths:}
        \begin{align}
            \sigma_{all_v}^{(0)} &= \mathcal{P}(A_v^L, \sigma_v, T^{(0)}), \\
            \sigma_{all_w}^{(0)} &= \mathcal{P}(A_w^L, \sigma_w, T^{(0)}), \\
            \sigma_{all}^{(0)} &= \sqrt{\sigma_{all_v}^{(0)} \sigma_{all_w}^{(0)}}.
        \end{align}

        \item \textit{Estimate the wake deficit using:}
        \begin{align}
            s_{D_c} &= \sqrt{2\ln 2} \, \sigma_{all}^0, \quad \xi = 1.1131, \\
            \beta &= \frac{\xi^2}{2(\sigma_{all}^{(0)})^2}, \quad a = \frac{1}{2\sigma_{all}^{(0)}\sqrt{2}}, \\
            \alpha_x^{(0)} &= \frac{(1 - \sqrt{1 - C_T})U_\infty\, \mathrm{erf}\left(\frac{s_{D_c}}{\sigma_{all}^{(0)}\sqrt{2}}\right)}
            {\sqrt{ \sigma_{all}^{(0)} \sqrt{2} \left[2a\,\mathrm{erf}(\sqrt{2}a) + \frac{\sqrt{2}}{\sqrt{\pi}}e^{-2a^2} \right] - \frac{1}{2} \sqrt{\frac{\pi}{\beta}} }}.
        \end{align}

        \item \textit{Estimate wake convective velocity:}
        \begin{equation}
            U_{c}^{(0)} = U_\infty - \frac{1}{2}\alpha_x^{(0)}.
        \end{equation}

        \item \textit{(Optional) Iterate until convergence:} \\
        For each iteration \( n \geq 1 \), update:
        \begin{align}
            T^{(n)} &= \frac{x - x_0}{U_{c}^{(n-1)}}, \\
            \sigma_{all_v}^{(n)} &= \mathcal{P}(A_v^L, \sigma_v, T^{(n)}), \quad
            \sigma_{all_w}^{(n)} = \mathcal{P}(A_w^L, \sigma_w, T^{(n)}), \\
            \sigma_{all}^{(n)} &= \sqrt{\sigma_{all_v}^{(n)} \sigma_{all_w}^{(n)}}, \\
            a^{(n)} &= \frac{1}{2\sigma_{all}^{(n)}\sqrt{2}}, \quad \beta^{(n)} = \frac{\xi^2}{2(\sigma_{all}^{(n)})^2}, \\
            \alpha_x^{(n)} &= \frac{(1 - \sqrt{1 - C_T})U_\infty\, \mathrm{erf}\left(\frac{s_{D_c}}{\sigma_{all}^{(n)}\sqrt{2}}\right)}
            {\sqrt{ \sigma_{all}^{(n)} \sqrt{2} \left[2a^{(n)}\,\mathrm{erf}(\sqrt{2}a^{(n)}) + \frac{\sqrt{2}}{\sqrt{\pi}}e^{-2(a^{(n)})^2} \right] - \frac{1}{2} \sqrt{\frac{\pi}{\beta^{(n)}}} }}, \\
            U_{c}^{(n)} &= U_\infty - \frac{1}{2} \alpha_x^{(n)}.
        \end{align}
        
        The iteration is repeated until $ \left|U_{c}^{(n)} - U_{c}^{(n-1)}\right| < \varepsilon $, where $\varepsilon $ is a prescribed tolerance. This step is optional, as the results obtained without iteration are already satisfactory and further iterations provide only marginal improvements.\\
    \end{enumerate}
    \item \textbf{Estimate the wake velocity deficit:}\\
    Finally, the wind velocity can be expressed as:
    \begin{equation}
       U(x,s_D) = \alpha_x^{(n)} \frac{1}{2} \left[\text{erf}\left(\frac{s_D + 0.5}{\sqrt{2}\sigma_{all}^{(n)}(x)}\right) - \text{erf}\left(\frac{s_D - 0.5}{\sqrt{2}\sigma_{all}^{(n)}(x)}\right)\right]
    \end{equation}

\end{enumerate}

\end{document}